\documentclass[prx, twocolumn,longbibliography]{revtex4-1}

\usepackage{float}
\usepackage{bbm}
\usepackage{epsfig}
\usepackage{epstopdf}
\usepackage{graphicx, subfigure}
\usepackage{amsmath,amssymb}
\usepackage{color}
\usepackage{hyperref}
\usepackage[capitalize]{cleveref}
\usepackage{braket,mleftright}

\usepackage{dsfont}
\usepackage{extarrows}
\usepackage{siunitx}
\usepackage{physics}

\begin{document}
\newcommand\Sx{S_x}
\newcommand\Sy{S_y}
\newcommand\Sz{S_z}
\newcommand\Ix{I_x}
\newcommand\Iy{I_y}
\newcommand\Iz{I_z}
\newcommand\sigx{\sigma_x}
\newcommand\sigy{\sigma_y}
\newcommand\sigz{\sigma_z}
\newcommand\gdd{|g \downarrow \Downarrow \rangle}
\newcommand\gdu{|g \downarrow \Uparrow \rangle}
\newcommand\gud{|g \uparrow \Downarrow \rangle}
\newcommand\guu{|g \uparrow \Uparrow \rangle}
\newcommand\edd{|e \downarrow \Downarrow \rangle}
\newcommand\edu{|e \downarrow \Uparrow \rangle}
\newcommand\eud{|e \uparrow \Downarrow \rangle}
\newcommand\euu{|e \uparrow \Uparrow \rangle}
\newcommand\down{|\Tilde{\Downarrow}\rangle}
\newcommand\up{|\Tilde{\Uparrow}\rangle}
\newcommand\downBra{\langle\Tilde{\Downarrow}|}
\newcommand\upBra{\langle\Tilde{\Uparrow}|}
\newcommand\downdown{|\Tilde{\Downarrow}\Tilde{\Downarrow}\rangle}
\newcommand\downup{|\Tilde{\Downarrow}\Tilde{\Uparrow}\rangle}
\newcommand\updown{|\Tilde{\Uparrow}\Tilde{\Downarrow}\rangle}
\newcommand\upup{|\Tilde{\Uparrow}\Tilde{\Uparrow}\rangle}
\newcommand\downdownBra{\langle\Tilde{\Downarrow}\Tilde{\Downarrow}|}
\newcommand\downupBra{\langle\Tilde{\Downarrow}\Tilde{\Uparrow}|}
\newcommand\updownBra{\langle\Tilde{\Uparrow}\Tilde{\Downarrow}|}
\newcommand\upupBra{\langle\Tilde{\Uparrow}\Tilde{\Uparrow}|}
\newcommand\DE{\Delta \! E}
\newcommand\dE{\delta \! E}

\title{Fast noise-resistant control of donor nuclear spin qubits in silicon}
\author{James Simon$^{1,2}$}
\author{F. A. Calderon-Vargas$^2$}
\author{Edwin~Barnes$^2$}
\author{Sophia~E.~Economou$^2$}
\email{economou@vt.edu}
\affiliation{$^1$Department of Physics, University of California, Berkeley, California 94720, USA\\
$^2$Department of Physics, Virginia Tech, Blacksburg, Virginia 24061, USA
}

\begin{abstract}
A high degree of controllability and long coherence time make the nuclear spin of a phosphorus donor in isotopically purified silicon a promising candidate for a quantum bit. However, long-distance two-qubit coupling and fast, robust gates remain outstanding challenges for these systems. Here, following recent proposals for long-distance coupling via dipole-dipole interactions, we present a simple method to implement fast, high-fidelity arbitrary single- and two-qubit gates in the absence of charge noise. Moreover, we provide a method to make the single-qubit gates robust to moderate levels of charge noise to well within an error bound of $10^{-3}$. 

\end{abstract}
\maketitle

\section{Introduction} \label{intro}
Nuclear spins in the solid state present unparalleled advantages as a platform for quantum computing due to their long coherence times~\cite{Saeedi2013,Muhonen2014} and high degree of controllability~\cite{Vandersypen2004a,Jones2010,Asaad2019}.
In particular, the nuclear spin of a phosphorus donor in silicon is a promising candidate for a quantum bit~\cite{Kane1998,Zwanenburg2013} owing to its coherent control~\cite{Pla2013,Muhonen2015,Muhonen2018} and minute-long coherence time~\cite{Muhonen2014}. The use of isotopically-purified silicon nanostructures~\cite{Itoh2014} considerably reduces magnetic environmental noise, allowing high-fidelity control~\cite{Pla2013}. 
However, long coherence times are only useful when gates are very fast in comparison.  One of the difficulties of using the nuclear spin as a qubit has been the implementation of fast single- and two-qubit gates. Controlling a nuclear spin with an oscillating magnetic field as in nuclear magnetic resonance is slow, with typical gate times ranging from a few to tens of microseconds~\cite{Pla2013,Muhonen2015,Sigillito2017b}. Moreover, most of the approaches for multi-qubit operations require short interaction distances~\cite{Kane1998}\cite{Hill2015}, and thus demand near-atomic precision in the placement of the donors \cite{Hile2018,He2019}. 

To overcome these challenges, Ref.~\onlinecite{Tosi2017} proposes the dynamical creation of a strong electric dipole transition at microwave frequencies for the nuclear spin by sharing an electron between the donor and the Si/SiO$_2$ interface and applying an oscillating magnetic field. This facilitates the implementation of two-qubit gates via dipole-dipole interactions or, alternatively, the qubit's coupling to other quantum systems. Moreover, nuclear spin-flip transitions can also be sped up by including an oscillating electric field along with the magnetic drive. A potential downside of making the system amenable to electrical control is that it increases its sensitivity to charge noise due to the charge component in the states encoding the qubit. However, Ref.~\onlinecite{Tosi2017} shows that there are regions in parameter space (``clock transitions'') where the nuclear spin transition is insensitive to electrical noise to at least first order.

In this work, we propose an alternative path toward robust high-fidelity single-qubit gates that does not rely on clock transitions. Our approach is based on using optimally designed control pulse waveforms and energy transition modulation. Accordingly, we derive a time-independent analytical approximation for the system's Hamiltonian, explain how to rapidly implement arbitrary, noise-resistant single-qubit gates with fidelities exceeding 99.9\% even in the presence of significant charge noise, and provide a method to use the dipole-dipole interaction to implement \textsc{cphase} gates across a distance of $\SI{0.5}{\micro\meter}$.  The single-qubit and \textsc{cphase} gates have maximum durations of $\SI{500}{\nano\second}$ and $\SI{750}{\nano\second}$, respectively, with special cases such as single-qubit $Z$ gates being much faster ($<\SI{25}{\nano\second}$). An advantage of our protocol compared to prior work \cite{Tosi2017} is that it does not require finely tuning the system to a clock transition. This is not done at the expense of gate performance, and our gates are as robust but faster than those of Ref.~\onlinecite{Tosi2017}.

The paper is organized as follows.  In Sec.~\ref{systemSection} we introduce the system and its Hamiltonian.  In Sec.~\ref{derivationSection}, we derive an analytical time-independent Hamiltonian following two approaches. In Sec.~\ref{stateSection}, we define the qubit states and explain how to implement robust single-qubit gates, specifically arbitrary $Z$-rotations and $X$-rotations. We give a method for implementing fast controlled-phase gates between two adjacent qubits separated by a distance of $\SI{0.5}{\micro\meter}$ in Sec.~\ref{2Qsection}.  We conclude in Sec.~\ref{conclusion}.

\section{The System} \label{systemSection}
\begin{figure}
\includegraphics[width=0.7\linewidth]{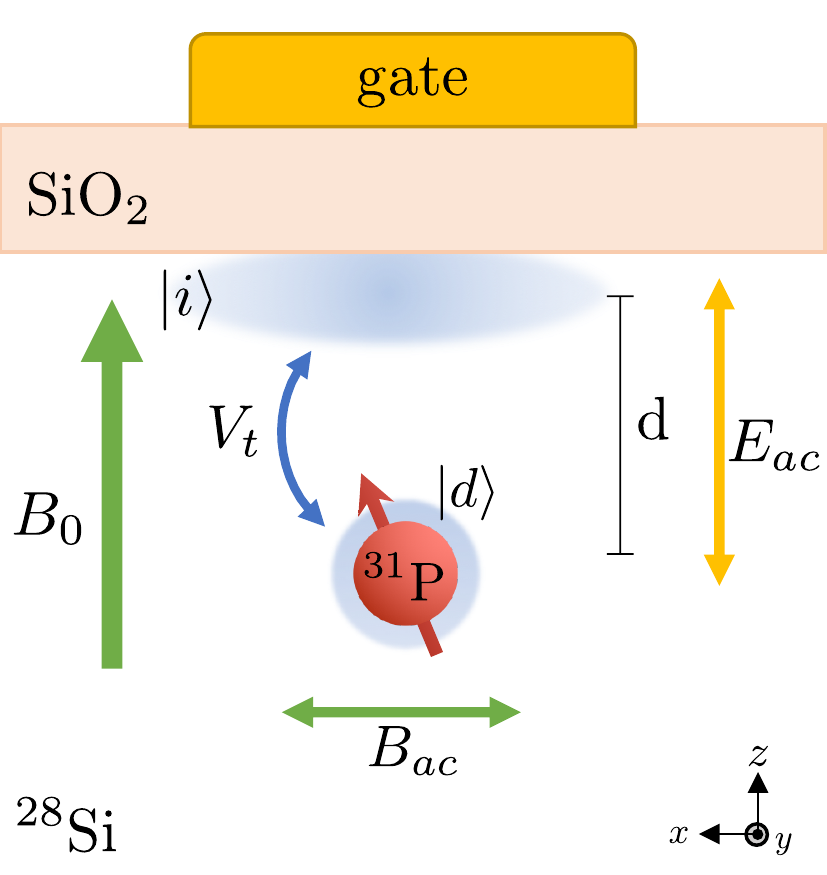}
\caption{One nuclear spin qubit in the system described.  A ${}^{31}$P donor is embedded in ${}^{28}$Si, and the free electron in the system can be pulled towards a Si-SiO${}_2$ interface.  The electron orbit is quantized into a $|d\rangle$ state on the donor and a $|i\rangle$ state on the interface, and both the electron and nuclear spins are used.  Static and oscillating electric and magnetic fields are used to control the system.  The qubit is ultimately stored in the nuclear spin state, with the other degrees of freedom used for driving gates.}
\label{systemSchematic}
\end{figure}

The system follows the experimental proposal reported in Refs.~\onlinecite{Tosi2017b,Tosi2017}, where a donor ${}^{31}\text{P}$ atom is embedded in enriched ${}^{28}\text{Si}$ a distance $d$ away from a $\text{Si}/\text{SiO}_2$ interface, as shown in Figure \ref{systemSchematic}. The donor atom provides a nuclear spin $I=1/2$ with gyromagnetic ratio $\gamma_n/2\pi=17.23 \text{ MHzT}^{-1}$ and a free electron with spin $S=1/2$ and gyromagnetic ratio $\gamma_e/2\pi= 27.97\text{ GHzT}^{-1}$. The electron and nuclear spins are coupled via a hyperfine interaction with coupling strength $A$, which is approximately equal to $\SI{117}{\MHz}$ when the electron is bound to the nucleus. A metal gate positioned on top of the donor atom is used to control the position of the electron via electric fields, which also tunes the hyperfine interaction $A$ down to zero when the electron is at the interface. Moreover, the gyromagnetic ratio of the electron bound to the nucleus can differ from that of an electron at the interface by an amount $\Delta\gamma$ that can reach up to 0.7\%~\cite{Freeman2016}.  Therefore, the Hilbert space includes three binary degrees of freedom: the nuclear spin of the donor atom, the spin of the free electron, and the position of the free electron, which is quantized into a state on the donor atom, $\ket{d}$, and a state at the interface, $\ket{i}$, which is a good approximation as demonstrated by Ref.~\onlinecite{Tosi2017b}. 

The system has two control fields, one electric and one magnetic, each with static (DC) and oscillating (AC) components. The DC component of the electric field, $E$, points along the donor-interface axis and controls the electron position (see Fig.~\ref{systemSchematic}), determining the amplitudes of the states $\ket{i}$ and $\ket{d}$ in the orbital ground state. The AC electric field is parallel to $E$ and is given by $E_{ac}(t) = E_{a}\cos(\omega_E t)$. In particular, when $E=E_0$, where $E_0$ is the magnitude of the electric field at the ionization point, the orbital ground state of the electron has equal probability to be at the donor nucleus and at the interface. When $E \ll E_0$, the electron is fully on the donor ($\ket{d}$), and when $E \gg E_0$ the electron is pulled off the donor ($\ket{i}$). The transition frequency between the orbital ground and excited states at the ionization point is equal to the tunnel coupling $V_t$. A strong static magnetic field $B_0$ ($B_0(\gamma_e+\gamma_n)\gg A$) splits the energy of the nuclear and electron spin states ($\{\ket{\Uparrow},\ket{\Downarrow}\}$ and $\{\ket{\uparrow},\ket{\downarrow}\}$, respectively). The AC magnetic field is perpendicular to $B_0$ and is given by $B_{ac} = B_{a}\cos(\omega_B t)$. The static electric and magnetic fields are parallel to avoid reductions in spin relaxation times caused by spin orbit effects ~\cite{Weber2018}. Note also that all-electrical spin control is possible even in the presence of a significant spin-orbit interaction, but it is vulnerable to charge noise ~\cite{Boross2018}. 

The Hamiltonian of the system, therefore, consists of the orbital part, the Zeeman part, and the hyperfine coupling:
\begin{equation}\label{eq:Total_Hamiltonian}
H = H_{orb} + H_{B} + H_A.
\end{equation}
Here each term can be expressed in terms of the electron position operators  $\tau^{id}_z = \ket{i}\bra{i}-\ket{d}\bra{d}$, $\tau^{id}_x = \ket{i}\bra{d}+\ket{d}\bra{i}$, and the electron (nuclear) spin operator $\mathbf{S}$ ($\mathbf{I}$) as follows:
\begin{equation}\label{eq:Hamiltonian_elements_position_basis}
\begin{aligned}
H_{orb} =& -\frac{d e \left(\DE +E_{a} \cos\left[\omega_E t \right]\right)}{2 \hbar} \tau^{id}_z + \frac{V_t}{2}\tau^{id}_x,\\
H_B =& B_0 \left( \gamma_e \left[ \mathds{1} + \left(\frac{\mathds{1} - \tau^{id}_z}{2}\right)\Delta\gamma \right] S_z - \gamma_n I_z \right)\\
& +B_{a}\cos\left[\omega_B t\right]\left(\gamma_e S_x - \gamma_n I_x\right),\\
H_A =& A \left(\frac{\mathds{1} - \tau^{id}_z}{2}\right) \mathbf{S} \cdot \mathbf{I},
\end{aligned}
\end{equation}
where $\Delta E=E-E_0$ is the deviation of the electric field away from the ionization point.

The qubit is encoded in the two lowest-energy eigenstates of the system, which, in the absence of AC driving, are approximately $\gdu$ and $\gdd$, where $\ket{g}$ is the ground eigenstate of the orbital part of the Hamiltonian with no AC fields. Therefore, it is convenient to express the total Hamiltonian in \eqref{eq:Total_Hamiltonian} in a basis spanned by the orbital eigenstates $\{\ket{g},\ket{e}\}$. The electron position operators $\tau_x^{id}$ and $\tau_z^{id}$ in the orbital eigenbasis are
\begin{equation}
\begin{aligned}
\tau^{id}_z = \frac{d e \DE}{\hbar \varepsilon_0} \tau_z + \frac{V_t}{\varepsilon_0} \tau_x,\\
\tau^{id}_x = - \frac{V_t}{\varepsilon_0} \tau_z + \frac{d e \DE}{\hbar \varepsilon_0} \tau_x,
\end{aligned}
\end{equation}
where $\tau_z=\ket{g}\bra{g}-\ket{e}\bra{e}$ and $\tau_x=\ket{g}\bra{e}+\ket{e}\bra{g}$ are the orbital operators, and $\varepsilon_0=\sqrt{V_t^2+(d e \DE/\hbar)^2}$ is the orbital (charge) energy splitting. The Hamiltonian components in Eq.~\ref{eq:Hamiltonian_elements_position_basis} have the following form in the basis spanned by the orbital and spin eigenbasis:
\begin{equation}
\begin{aligned}
H_{orb}=&\frac{-\varepsilon_0}{2}\tau_z-\frac{d e E_{a} \cos(\omega_E t)}{2 \hbar}\left( \frac{d e \DE}{\hbar \varepsilon_0}\tau_z + \frac{V_t}{\varepsilon_0} \tau_x \right),\\
H_{B} =& B_0 \gamma_e \left[ \mathds{1} + \left(\frac{\mathds{1}}{2} - \frac{d e \DE / \hbar \tau_z + V_t \tau_x}{2 \varepsilon_0}\right)\Delta\gamma \right] S_z\\
& -B_0\gamma_n I_z +B_{a}\cos\left[\omega_B t\right]\left(\gamma_e S_x - \gamma_n I_x\right),\\
H_A =& A \left( \frac{\mathds{1}}{2} - \frac{d e \DE}{2 \hbar \varepsilon_0} \tau_z -  \frac{V_t}{2 \varepsilon_0} \tau_x \right) \mathbf{S} \cdot \mathbf{I}.
\end{aligned}
\end{equation}

\section{Deriving the Time-Independent Hamiltonian} \label{derivationSection}

The dominant energy scales of this system are the charge splitting $\varepsilon_0$ and the electron spin splitting $B_0 \gamma_e$, which are driven at frequencies $\omega_E$ and $\omega_B$, respectively.  We transfom into a rotating frame involving both frequencies, leaving a Hamiltonian that is largely static. Therefore, we move the Hamiltonian to the rotating frame $\Tilde{H} = \Lambda H \Lambda^\dagger - i \Lambda \dot{\Lambda}^\dagger$ with $\Lambda =\exp\left[-i t \left(\omega_E (\tau_z/2 + I_z) - \omega_B (S_z+I_z) \right) \right]$, where the system's dominant off-diagonal energy terms ($B_{a}\gamma_e$, $\frac{E_{a} d e}{\hbar}$ and $\tfrac{A V_t}{\varepsilon_0}$) become effectively static. Assuming that the driving fields' detunings are small and the orbital and electron spin splittings are similar (i.e. $\varepsilon_0\approx\omega_E\approx B_0\gamma_e\approx\omega_B$), we can then apply the rotating wave approximation (RWA), dropping all rapidly oscillating terms from $\Tilde{H}$ to get:
\begin{equation} \label{RWAHamiltonianEquation}
\begin{aligned}
\Tilde{H}_0 &= \frac{-\varepsilon_0 + \omega_E}{2}\tau_z - E_{a}(t) \frac{d e V_t}{4 \hbar \varepsilon_0} \tau_x + (B_0 \gamma_e - \omega_B) S_z\\
&- (B_0 \gamma_n + \omega_B - \omega_E) I_z + B_0 \gamma_e \Delta\gamma \left(\frac{\mathds{1}}{2} - \frac{d e \DE \tau_z}{2 \hbar \varepsilon_0} \right)S_z \\
&+ \frac{B_{a}(t) \gamma_e}{2} S_x  + \frac{A}{2}\left(\mathds{1} - \frac{d e \DE}{\hbar \varepsilon_0} \tau_z \right)S_z I_z \\
&- \frac{A V_t}{4 \varepsilon_0} \left( \ket{g \uparrow \Downarrow} \bra{ e \downarrow \Uparrow} + \ket{e \downarrow \Uparrow}\bra{g \uparrow \Downarrow} \right).
\end{aligned}
\end{equation}
Except for the final term coupling the $\gud$ and $\edu$ states, this Hamiltonian is entirely diagonal when the driving fields are zero.

This approximate RWA Hamiltonian works very well for short times.  However, as the system evolves, the approximate evolution gradually becomes dephased relative to the true evolution due to energy shifts caused by the dropped high-frequency terms, in a similar effect to the Bloch-Siegert shift \cite{BlochSiegert}. Corrections to the RWA are required, and thus we use multi-frequency Floquet theory to construct a Floquet Hamiltonian $H_F$ that takes into account these higher-frequency modes. Then we use 2nd-order quasi-degenerate perturbation theory to reduce it to a (still non-oscillating) approximation $H'$ that reproduces the evolution given by the oscillating lab-frame Hamiltonian with typical fidelity $> 0.9999$ for the gates we describe in this paper.  The derivation of this approximation is detailed in Appendix~\ref{AFloquet}.

As with $\Tilde{H}_0$, this approximation's only time dependence is in the changing envelopes of the control pulses, making it both conceptually simpler and faster to simulate with common software.  It is thus useful for optimizing gates or running accurate simulations much faster than with the lab-frame Hamiltonian.  In this paper, we use the lab-frame Hamiltonian in all of our final results for the sake of caution, but we have confirmed that using this approximation gives the same results.

\section{Qubit states and gates} \label{stateSection}

The qubit is defined to be the two lowest-energy eigenstates of the system. In the absence of driving, the electron is predominantly in the ground orbital with spin down, so these states are approximately equal to $\gdu$ and $\gdd$.  We refer to the exact qubit states as $\up$ and $\down$, respectively.  With zero AC fields and typical parameter values (given below), the approximations $\up \approx \gdu$ and $\down \approx \gdd$ hold to within an overlap error of at most $10^{-4}$, which becomes much lower still when $\DE > 0$.

\begin{figure}
\includegraphics[width=\linewidth]{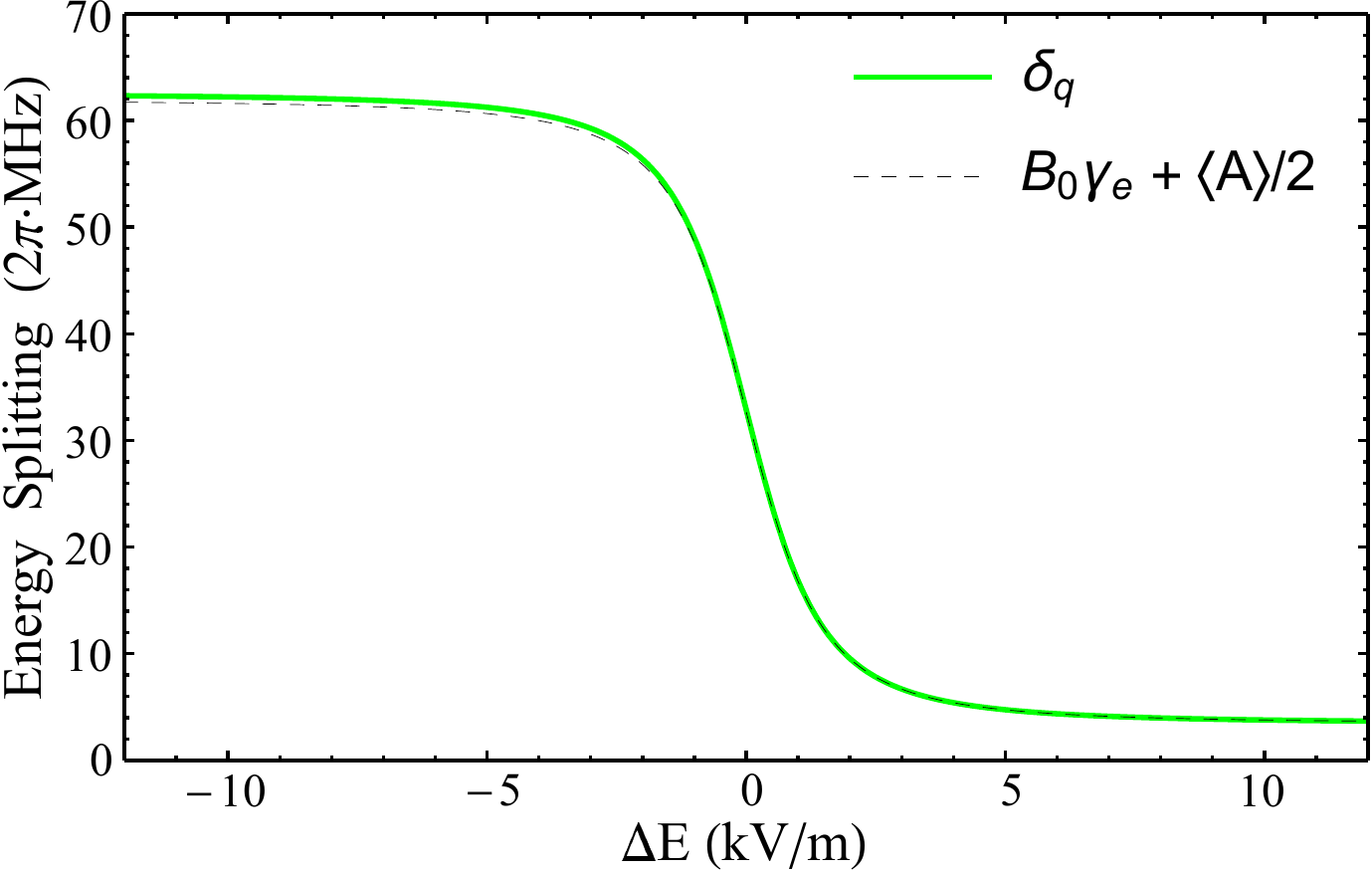}
\caption{Energy splitting $\delta_q$ between qubit states in the lab frame as a function of $\DE$ with no driving fields.  Both the numerical value and the approximation given in Eq.~\ref{deltaQApprox} are shown.  Varying $\DE$ changes this splitting, allowing the implementation of $R_z(\theta)$ gates.} \label{dephasingPlot}
\end{figure}

The dominant source of decoherence in this system is quasi-static charge noise with a typical $1/f$ spectrum~\cite{Paladino2014,Freeman2016}. We model the quasi-static charge noise as acting along the $z$-axis, directly perturbing the applied electric field $\DE$~\cite{Note_Lateral_Noise}. The energy splitting $\delta_q$ between the two qubit states depends on the applied electric field (see Figure \ref{dephasingPlot}), so noise in the electric field will lead to uncertainty in the energy difference of the qubit states, causing dephasing.  This energy splitting can be very accurately approximated by
\begin{equation} \label{deltaQApprox}
    \delta_q \equiv E_{\Tilde{\Downarrow}} - E_{\Tilde{\Uparrow}} \approx B_0 \gamma_n + \frac{\langle A \rangle}{2},
\end{equation}
where $E_{\Tilde{\Downarrow}}$ and $E_{\Tilde{\Uparrow}}$ are the energies of the respective states and $\langle A \rangle = A \rvert \langle g | d \rangle \rvert^2 = (A/2)(1 - d e \DE / \varepsilon_0)$.  If the system idles for a time $t$ with a small error in the electric field $\dE$, we expect the dephasing to be approximately given by $R_z\Big(-\frac{d \delta_q}{d \DE} \dE t\Big)$.

Following Ref.~\onlinecite{Tosi2017}, in all the following calculations we use $A/2\pi =\SI{117}{\MHz}$, $\gamma_e/2\pi =\SI{27.97}{\GHz\per\tesla}$, $\gamma_n/2\pi =\SI{17.23}{\MHz\per\tesla}$,  and $\Delta\gamma = -.002$, and choose $d = \SI{15}{\nm}$, $B_0 = \SI{.2}{\tesla}$, and $V_t = B_0 (\gamma_e + \gamma_n)$. Moreover, we choose $\DE = \SI{e4}{\volt\per\meter}$ so that in the absence of AC driving, the qubit idles with the electron at the interface, in a region where $d \delta_q/d \DE$ is very small and dephasing is thus reduced (see Appendix \ref{Adephasing} for further discussion). Every gate shown here starts and ends at this chosen idling point. We also define our gates in the idling frame of the qubit, so the evolution while idling is simply the identity.  

Any single-qubit gate can be decomposed into the form $R_z(\theta_{z1})R_x(\theta_x)R_z(\theta_{z2})$, where a $Z$ rotation is defined as $R_z(\theta) = \exp\left[ -i \tfrac{\theta}{2}\sigma_z \right]$, $\theta \in [0, 2\pi)$, and an $X$ rotation is defined as $R_x(\theta) = \exp\left[ -i \tfrac{\theta}{2}\sigma_x \right]$, $\theta \in [0, \pi)$.  The qubit Pauli operators are defined following $\sigma_z = \up\upBra - \down\downBra$. Below we show how to implement a noise-resistant $R_z(\theta)$ gate.

\subsection{$\mathbf{R_z(\theta)}$ Gates} \label{Zsection}

Figure \ref{dephasingPlot} shows that simply changing $\DE$ changes $\delta_q$, causing a phase difference between the qubit states to accumulate. If leakage were not a concern, one could implement an effectively noiseless $Z$ rotation by shifting from the idling condition of $\DE \gg 0$ to $\DE \ll 0$.  $\delta_q$ then shifts by $\sim 2\pi \cdot \SI{60}{\mega\hertz}$, causing a full $2\pi$ rotation in $\sim\SI{20}{\nano\second}$, and $d \delta_q / d \DE$ will be small as during idling. However, the non-oscillating electric field $\DE$ cannot be shifted arbitrarily fast, and changing $\DE$ changes the transformation we have used in going from the $\{\ket{i}, \ket{d}\}$ basis to the $\{\ket{g}, \ket{e}\}$ basis, causing another term to appear in $H$ (as derived in Appendix \ref{Age}) that can drive the system out of the logical space. Nonetheless, the effect on the fidelity is negligible for shift times of a few ns, so this does not significantly affect the gate time.  The resulting evolution will be adiabatic in the qubit subspace, so we only need to consider the phase accumulated.

We can calculate the phase accumulated using
\begin{equation} \label{phiZEquation}
    \theta = -\int^T_0 (\delta_q(t) - \delta^0_q) dt,
\end{equation}
where $\delta^0_q$ is the qubit splitting while idling, which we subtract so we work in a frame where, at idling, the evolution operator is the identity. For the pulse shapes, we use a cosine window function $w$ defined as follows:
\begin{equation}\label{eq:window_function}
w(t, \tau, T) = \begin{cases}
[1 - \cos(\pi t / \tau)]/2 & 0 \le t < \tau, \\
1 & \tau \le t < T - \tau, \\
[1 - \cos(\pi (T - t) / \tau)]/2 & T - \tau \le t \le T, \\
0 & t < 0 \text{ or } t > T.
   \end{cases}
\end{equation}

\begin{figure}
\includegraphics[width=\linewidth]{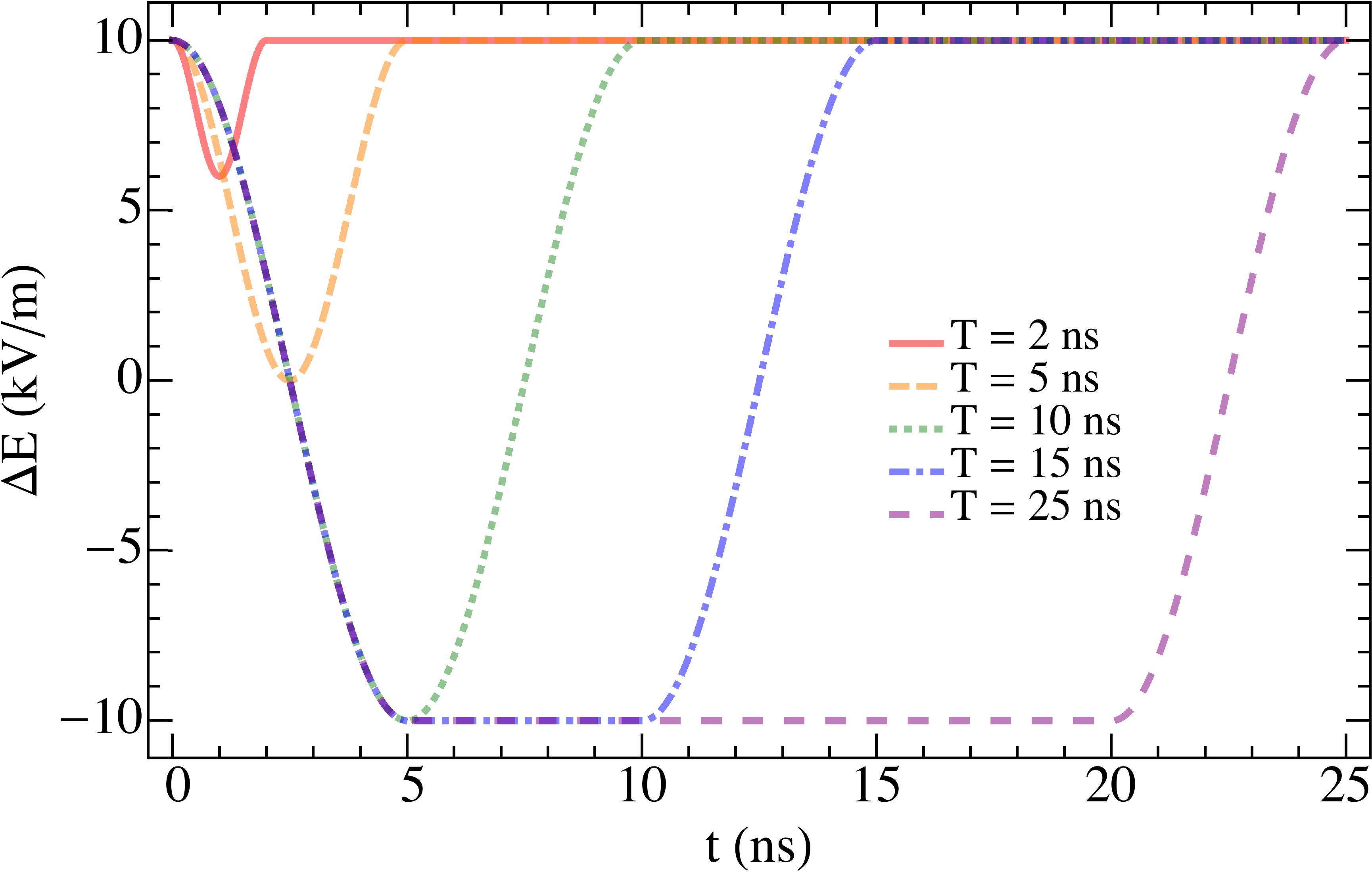}
\caption{Examples of $\DE$ pulses used to implement $R_z(\theta)$ gates for various values of the total gate time.  For long times, $\DE$ is moved to its minimum value, is held for several ns, and returns.  For short times, $\DE$ cannot be changed fast enough to reach the minimum while remaining adiabatic, so a shallower pulse is used.} \label{ZEpulsePlot}
\end{figure}

When implementing an $R_z(\theta)$ gate as described above, the varying function $\DE(t)$ is constrained by the fact that $\DE$ has to change slowly enough for evolution to be adiabatic.  We choose a minimum time of $\SI{5}{\nano\second}$ to move from idling to minimum $\DE$, so, for $R_z(\theta)$ gates with total time $T$ shorter than $\SI{10}{\nano\second}$, there will not be time for $\DE$ to reach its minimum value, so we simply make the $\DE(t)$ pulse shallower.  We choose $\DE(t) = \DE_{idle} - S \cdot w(t, \tau, T)$, where $E_{idle} = \SI{e4}{\volt\per\meter}$, $\tau = \min(\SI{5}{\nano\second}, T/2)$ and $S = (\SI{2e4}{\volt\per\meter}) \cdot \min(1, T/(\SI{10}{\nano\second}))$.  Figure \ref{ZEpulsePlot} shows several examples of $\DE(t)$ for various $T$.  As no AC fields are needed, we choose $E_{a}(t) = B_{a}(t) = 0$.

\begin{figure}
\includegraphics[width=\linewidth]{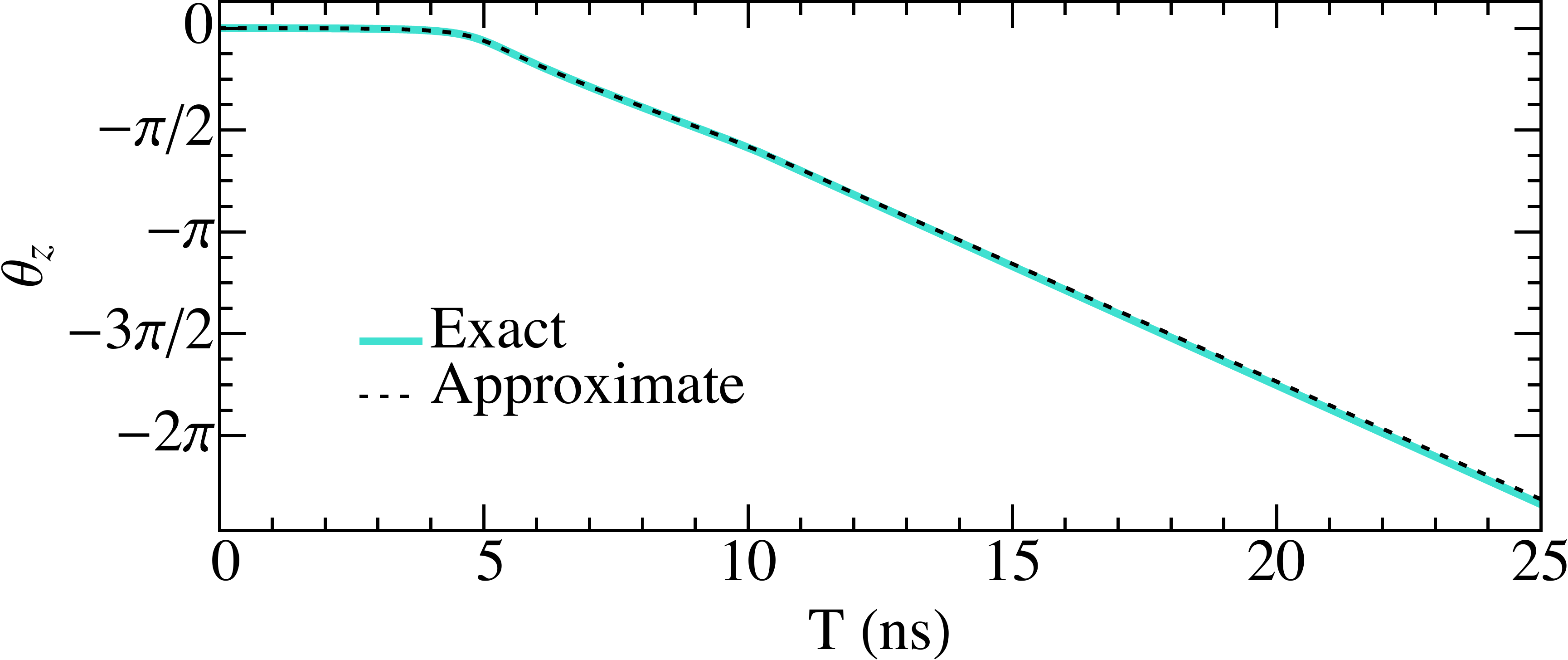}
\caption{Exact and approximate $R_z(\theta)$ angles for different gate durations $T$ using the gate scheme described in Section \ref{Zsection}.  The approximate curve is found from the approximation in Eq.~\ref{deltaQApprox}, integrated according to Eq.~\ref{phiZEquation}.  The exact and approximate curves agree to within an angle of .08 rad.} \label{phiZPlot}
\end{figure}

The angle of the implemented $R_z(\theta)$ gate as a function of $T$ is plotted in Fig.~\ref{phiZPlot}.  An arbitrary $Z$-rotation can be generated in under \SI{25}{\nano\second}, and a $Z$-gate with $\theta = \pi$ is generated in $T \approx \SI{14}{\nano\second}$.  The numerically calculated phase is quite close to a simple approximation obtained using Eqs.~\ref{deltaQApprox} and \ref{phiZEquation}, demonstrating that the phase of the gate is accumulating as we describe. While these gates are quite fast, we note that the 5 ns ramping time of our pulses is well within the risetime limitations of typical waveform generators. In the event that timing errors become an issue, these $R_z(\theta)$ gates could be combined with the $R_x(\theta)$ gates described below to construct a BB1 sequence \cite{Wimperis1994} to correct over-rotation errors.

\begin{figure}
\includegraphics[width=\linewidth]{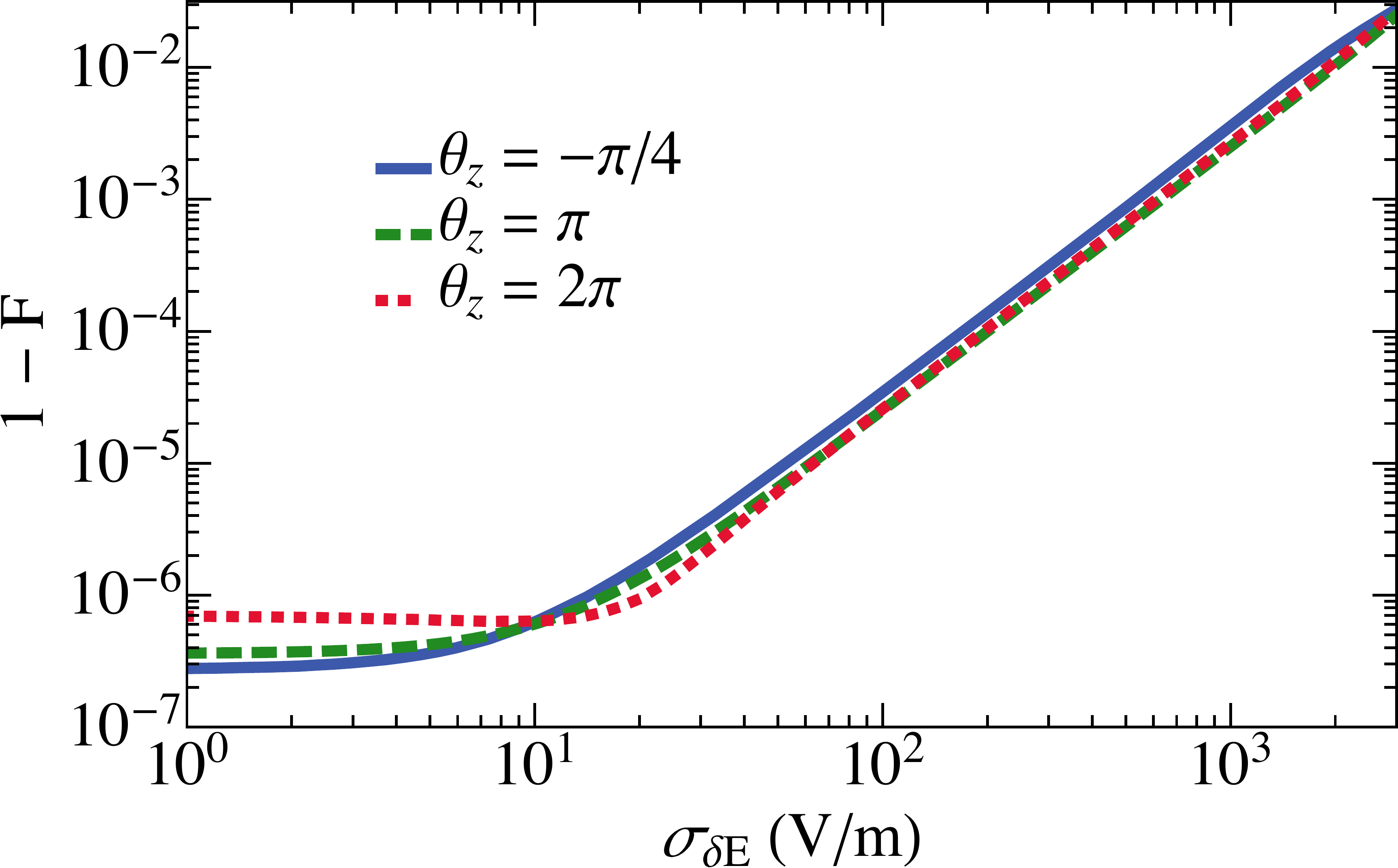}
\caption{Gate infidelity vs noise strength for three $R_z(\theta)$ gates with $\theta = -\pi/4$ ($T = \SI{6.632}{\nano\second}$), $\theta = \pi$ ($T = \SI{13.560}{\nano\second}$), and $\theta = 2\pi$ ($T = \SI{22.116}{\nano\second}$).  The numerical error in our simulations is roughly $10^{-6}$.} \label{ZnoisePlot}
\end{figure}

As mentioned before, charge noise is the most deleterious source of error for this type of system~\cite{Tosi2017b,Harvey-Collard2017a}. Although this noise has been measured to have a $1/f$ power spectrum \cite{Freeman2016}, this noise is largely concentrated at frequencies below 1 kHz. Because our gate times are several orders of magnitude faster than the noise fluctuation timescale, it should be a good approximation to treat the noise as quasi-static \cite{Martins2016}. In this work, we model this noise by adding a constant stochastic error $\delta E$ to $\DE$ for the duration of a gate. Since the charge noise is statistical in nature, we draw $\delta E$ from a normal distribution with standard deviation $\sigma_{\delta E}$ and report average infidelity over the distribution. The gate infidelity is defined as~\cite{Pedersen2007} $1-\mathcal{F}=1-\tfrac{1}{n(n+1)}[\Tr(U^{\dagger}U)+\vert\Tr(U_0^{\dagger}U)\vert^2]$, where $n$ is the Hilbert space dimension, $U$ is the generated gate, and $U_0$ is the desired gate. Figure \ref{ZnoisePlot} shows the gate infidelity for three different angles of rotation.  For typical noise with an r.m.s. of $\SI{100}{\volt\per\meter}$~\cite{Tosi2017b}, the error is well below $10^{-4}$ for all angles. Our $R_z(\theta)$ scheme is quite noise-resistant as it is, because the system spends most of its time with $\DE \ll 0$ or $\DE \gg 0$, so that $d \delta_q / d \DE$ is small for most of the gate.

\subsection{$\mathbf{R_x(\theta)}$ Gates} \label{Xsection}

\begin{figure}
\includegraphics[width=\linewidth]{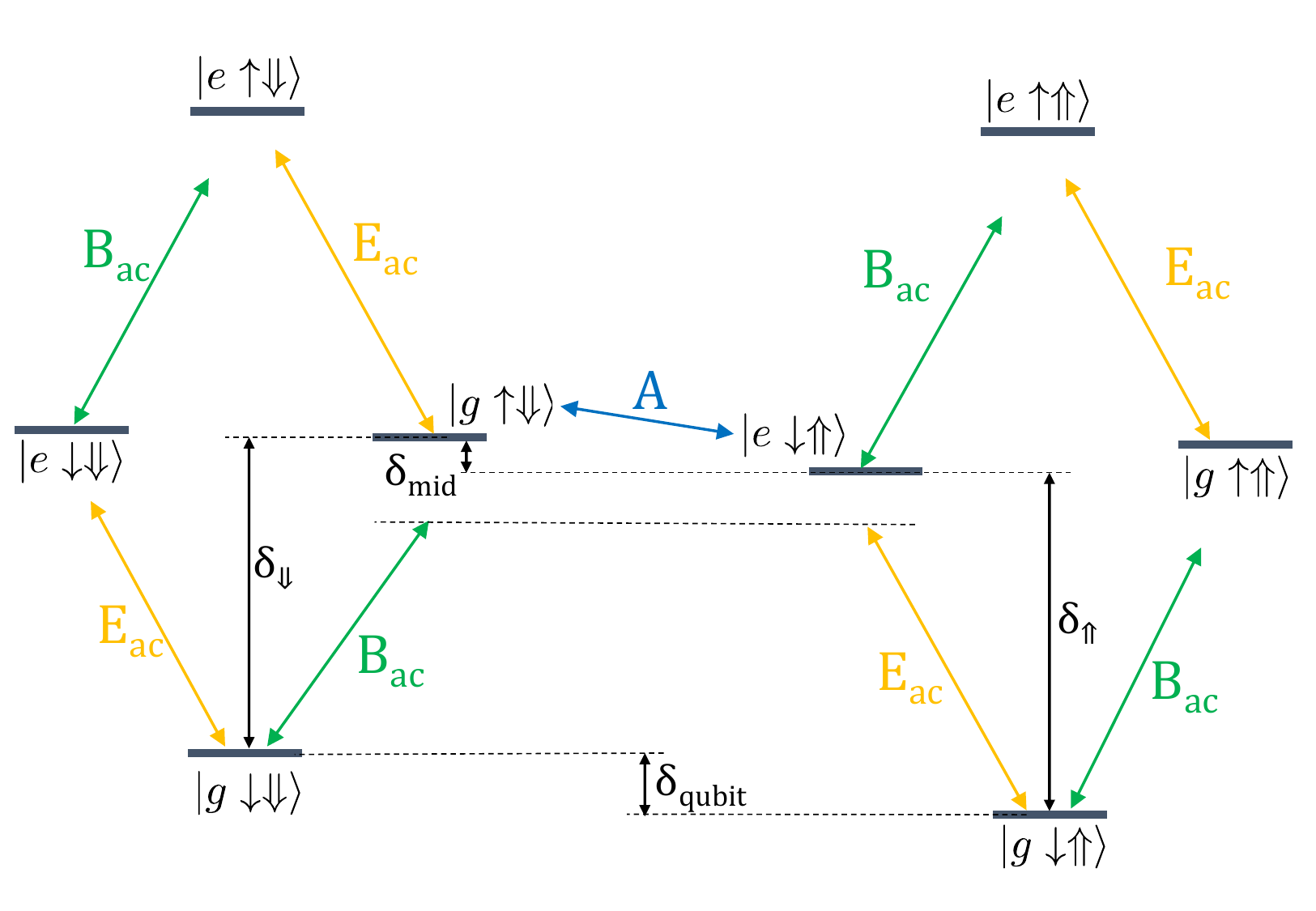}
\caption{A level diagram showing the states of the system in the original, lab basis and the significant off-diagonal terms (both oscillating and static).  While idling, the qubit states are the lowest two states.  There are other off-diagonal terms involving $A$ and $B_{a}$, but they are far off resonance and are of minor significance, contributing to correction terms in $H'$.} \label{allLevels}
\end{figure}

An $R_x(\theta)$ gate requires a coupling between the two qubit states.  The simplest way to achieve this is by driving $B_{ac}$ on resonance with the nuclear spin, but the scale of $\gamma_n$ makes this slow, with typical times ranging from a few to tens of microseconds for $\pi$ rotations~\cite{Pla2013,Muhonen2015,Sigillito2017b}.  A faster solution in this system is to create an effective coupling through intermediate non-computational states.  Figure \ref{allLevels} shows a level diagram of the system including all strong, near-resonance couplings between states, as reflected in $\tilde H_0$. The simplest way to drive fast transitions between the qubit states is to use two intermediate states:
\begin{equation*}
\gdd \xLongleftrightarrow{B_{ac}} \gud \xLongleftrightarrow{A} \edu \xLongleftrightarrow{E_{ac}} \gdu .
\end{equation*}

The first major problem one encounters is the fact that, if the driving frequencies are simply chosen to be on resonance, there will be severe leakage into the non-computational states.  The complexity of the system and the limited number of controllable parameters would seem to make this a difficult problem.  However, with slightly different choices of driving fields, one can drive a transition adiabatically with respect to the qubit subspace.  If $\DE$ and $B_0$ are chosen so that the lab-frame energies of $\gud$ and $\edu$ are similar (i.e., $\varepsilon_0 \approx B_0 (\gamma_e + \gamma_n)$), the hyperfine coupling strongly hybridizes these states.  Transitions between the qubit states can then be seen as similar to Raman transitions \cite{RamanTransition}, but with two intermediate states instead of one.  The driving field amplitudes and frequencies can be set like in an adiabatic Raman transition (i.e. making the driving matrix elements equal and the large detunings from intermediate states the same) to give a transition close to an $X$-gate with a leakage probability below $10^{-4}$.  The large detunings separate the qubit subspace from the rest of the Hilbert space, and we use slowly-varying pulse shapes instead of square pulses, so the resulting transition is very adiabatic.  The driving amplitudes and frequencies can then be optimized while worrying only about the computational subspace, while also using the $R_z(\theta)$ gates described previously to cancel unwanted accumulated phases, giving a high-fidelity $R_x(\theta)$ for any given angle.

This gate scheme, however, is extremely sensitive to electrical noise.  This can be seen by writing out the three relevant transition energies as follows:
\begin{equation}
\begin{aligned}
    \delta_\Downarrow &\equiv E_{\gud} - E_{\gdd} \approx B_0 \gamma_e - \langle A \rangle / 2,\\
    \delta_\Uparrow &\equiv E_{\edu} - E_{\gdu} \approx \varepsilon_0 - A/4 + \langle A \rangle/2,\\
    \delta_{mid} &\equiv E_{\edu} - E_{\gud} \approx \varepsilon_0 - B_0(\gamma_e + \gamma_n) - A/4 + \langle A \rangle / 2.
\end{aligned}    
\end{equation}
All three of these transition frequencies depend on $\DE$.  In addition, $\varepsilon_0$ strongly depends on $\DE$ when $|\DE| \gg 0$, while $\langle A \rangle$ strongly depends on $\DE$ when $\DE \approx 0$, so there is no value for $\DE$ that ameliorates this problem.  If any of the three transitions are far off resonance, the effective coupling between the qubit states approaches zero and an $X$-rotation becomes impossible.  Thus, while the two qubit states are effectively coupled, $\DE$ has to be known to high precision.

\begin{figure}
\includegraphics[width=\linewidth]{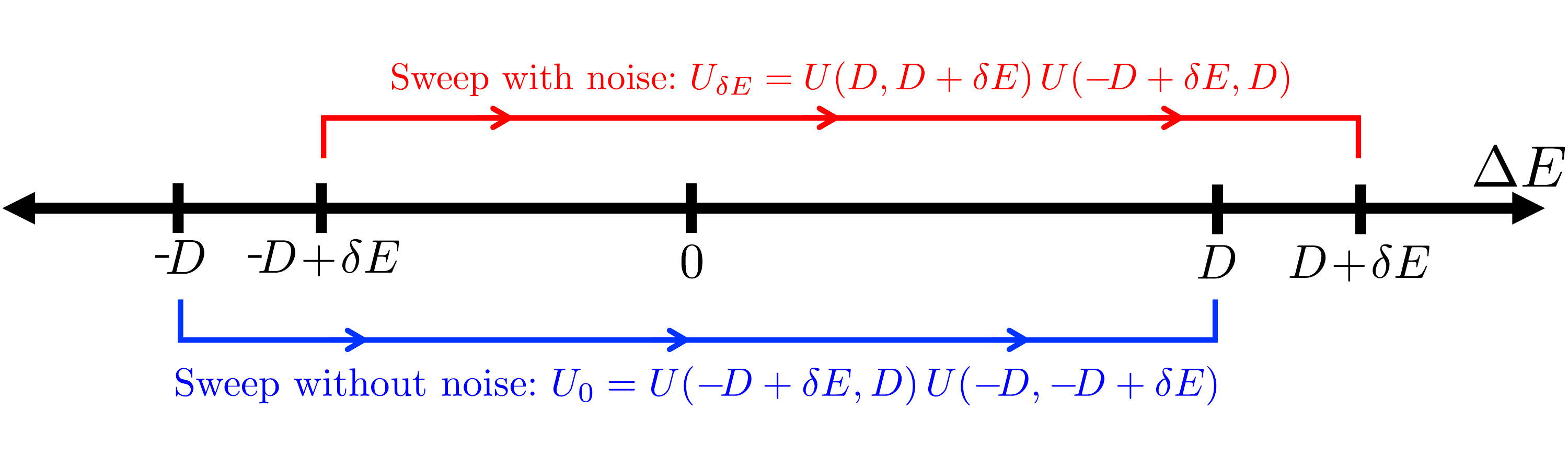}
\caption{The electric field $\DE$ during a sweep gate, with and without static error $\dE$ in the electric field.  The resulting evolution operators are shown.  As explained in the text, a key fact is that the middle stretch of the evolution, $U(-D + \dE, D)$, is the same with and without noise, leaving only small differences on the edges that constitute $R_z$ errors.} \label{sweepLineDiagram}
\end{figure}

Our solution to this is to have $\DE$ sweep through a broad range of values at a fixed rate instead of remaining constant.  The probability transfer between the two qubit states will happen in a short period in the middle whenever $\DE \approx 0$.  If there is quasi-static charge noise, the constant sweep rate ensures that this transition will still happen identically, but it will simply be shifted slightly in time, and there will be only noisy $Z$-rotations before and after the gate.  To explain this, consider sweeping $\DE$ from $-D$ to $D$ at a constant rate, first without error in $\DE$, and second with an error of $\dE > 0$, so $\DE$ actually sweeps from $-D + \dE$ to $D + \dE$.  In both cases, there is a segment of the evolution in which $\DE$ sweeps from $-D + \dE$ to $D$.  The difference is that in the first case, there is a sweep from $-D$ to $-D + \dE$ before this segment, and in the second case, there is a sweep from $D$ to $D + \dE$ after it.  Far from $\DE \approx 0$, the evolution operator will be diagonal in the qubit subspace, so these small pieces are just $R_z$ rotations, so the effect of this error $\dE$ is just to introduce phase errors before and after the gate. To explain why, consider a gate that sweeps from $\DE = -D$ to $\DE = D$ at a constant rate.  Let $U(a, b)$ represent the evolution operator resulting from sweeping $\DE$ through the range $[a, b]$. As shown in Fig.~\ref{sweepLineDiagram}, in the ideal case, we get the evolution $U_0 = U(-D + \dE, D) U(-D, -D + \dE)$, while if there is a static charge noise $\dE$ the actual evolution operator is $U_{\delta E} = U(D, D + \delta E) U(-D + \delta E, D)$.  The middle stretches of these evolution operators, from $\DE = -D + \dE$ to $\DE = D$, are essentially identical; apart from the ramping up of the AC fields, all parameters as a function of time are the same.  If $D$ is large, then the AC electric driving is very far off resonance at the ends of the evolution span, and there is no effective coupling between the computational states, so the error operators $U(D, D + \dE)$ and $U^\dagger(-D, -D + \dE)$ are entirely diagonal and amount to $R_z$ errors.  The effect of small charge noise, then, is only to add dephasing before and after the sweep gate.

As discussed above, an arbitrary gate $U$ can be decomposed into Euler angle form $U = R_z(\theta_{z1})R_x(\theta_x)R_z(\theta_{z2})$, and the three angles of this decomposition vary continuously with $U$ except when $\theta_x = 0, \pi$ due to a phenomenon called gimbal lock.  Except at these points, then, to first order, an arbitrary gate's dependence on charge noise can be decomposed as $R_z(\theta_{z1} + \theta_{z1}' \dE)R_x(\theta_x + \theta_x' \dE)R_z(\theta_{z2} + \theta_{z2}' \dE)$.  The effect of this sweep will be to make $\theta_x' \approx 0$, eliminating the noise-dependence of the $R_x$ component.  Additionally, tweaking the start and end points of the sweep can change $\theta_{z1}'$ and $\theta_{z2}'$.  In the special case of $\theta_x = \pi$, $\theta_{z1}' = \theta_{z2}'$, the two error terms cancel because $R_z(\theta_z)R_x(\pi) = R_x(\pi)R_z(-\theta_z)$, and the noise dependence is eliminated.  Gimbal lock is not a problem here because in practice $\theta_x \neq \pi$, so $\theta_{z1}'$ and $\theta_{z2}'$ do not diverge, and even at $\theta_x = \pi$ they diverge together and the divergences cancel. This means that to create an $X$-gate, we can apply a $R_z(-\theta_{z1} + \theta_{z2})$ gate after the sweep to remove the residual $R_z$ gates and leave only $R_x(\pi) \equiv X$. We use numerical simulations in conjunction with the Euler decomposition to determine the driving parameters necessary to produce this cancellation.

\begin{figure}
\includegraphics[width=\linewidth]{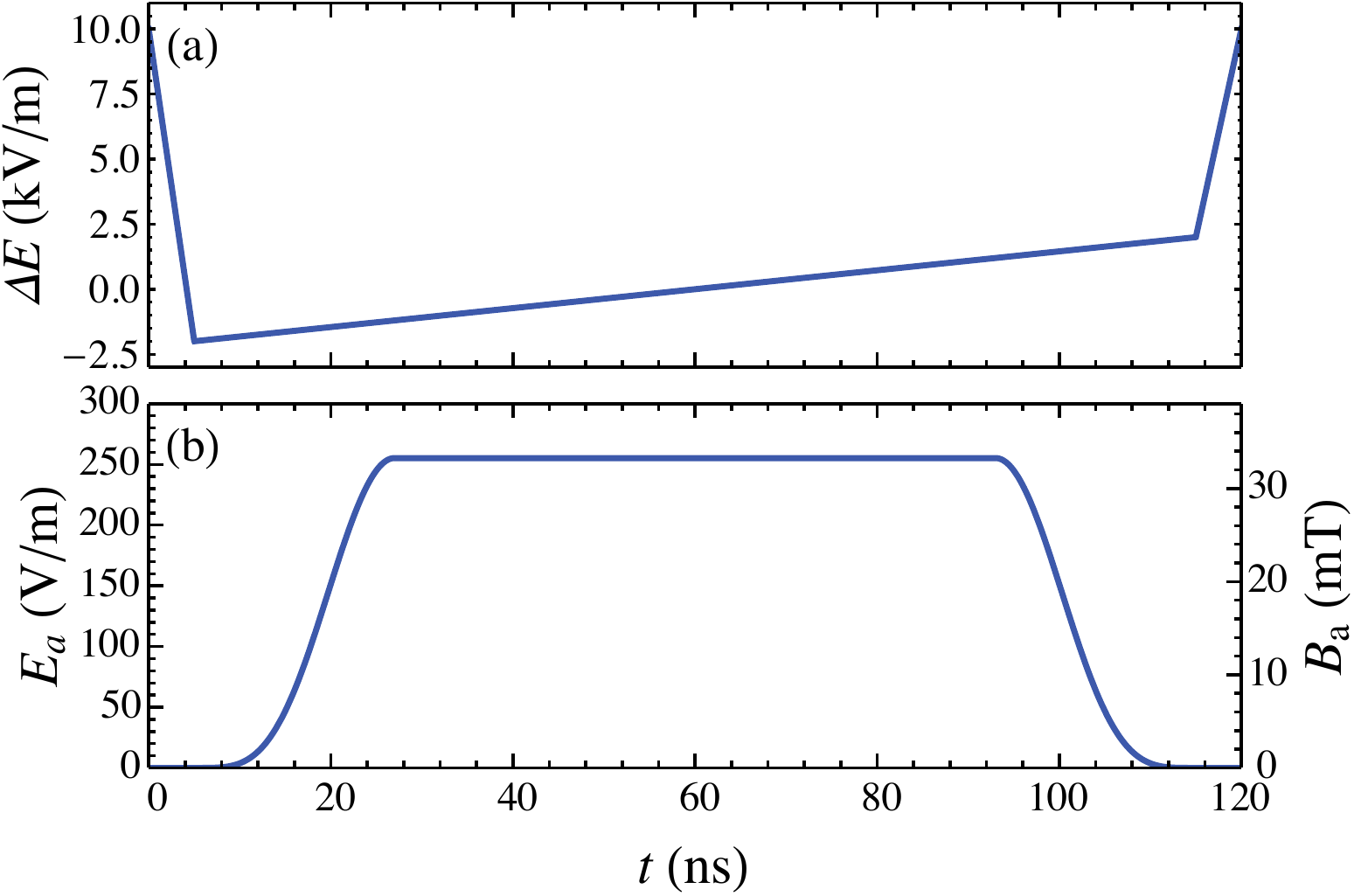}
\caption{Control pulses for $\DE$, $E_{a}$, and $B_{a}$ during a sweep gate giving a $R_x(\pi)$ rotation.  The AC fields only turn on during the middle $\SI{110}{\nano\second}$ when $\DE$ is steadily sweeping through zero.  An $R_z(\theta)$ gate (not shown) is also applied before or after to cancel extra phases accumulated during the sweep.} \label{XPulsePlots}
\end{figure}

Our precise control protocol for implementing a noise-resistant $X$-gate is summarized as follows. We sweep $\DE$ from $\SI{-2000}{\volt\per\meter}$ to $\SI{2000}{\volt\per\meter}$ in $\SI{110}{\nano\second}$ using the function $l$:
\begin{equation}
l(t, \tau_1, y_1, \tau_2, y_2, T) = \begin{cases}
y_1 t/\tau_1 & 0 \le t < \tau_1, \\
y_1 + (y_2 - y_1)\frac{t - \tau_1}{\tau_2 - \tau_1} & \tau_1 \le t < \tau_2, \\
y_2 \frac{T - t}{T - \tau_2} & \tau_2 \le t \le T, \\
0 & \hbox{otherwise}.
\end{cases}
\end{equation}
Then the control pulse is $\DE(t) = \DE_{idle} + l(t, \tau_1, -\DE_{idle} - D, \tau_1 + \tau_s, -\DE_{idle} + D, 2\tau_1 + \tau_s)$, where $\DE_{idle} = \SI{10000}{\volt\per\meter}$ is the idling voltage, $D = \SI{2000}{\volt\per\meter}$ is the amplitude of the sweep, $\tau_1 = \SI{5}{\nano\second}$ is the setup time and $\tau_{s} = \SI{110}{\nano\second}$ is the duration of the sweep.  For the AC driving fields, we use the window functions from Eq.~\ref{eq:window_function} and choose $E_{a}(t) = \lambda \cdot (\SI{255.2}{\volt\per\meter}) \cdot w^2(t - \tau_1, \tau_2/5, \tau_2)$, $B_{a}(t) = \lambda \cdot (\SI{33.26}{\milli\tesla}) \cdot w^2(t - \tau_1, \tau_2/5, \tau_2)$, where $\tau_2=\tau_1+\tau_s$ and $\omega_E = \varepsilon_0 - 2\pi \cdot \SI{232.428}{\mega\hertz}$, $\omega_B = B_0 \gamma_e - A/4 - 2\pi \cdot \SI{217.096}{\mega\hertz}$. Squaring the window function produces pulses that turn on more gradually, improving adiabaticity. We include the free parameter $\lambda$ to control $\theta_x$; the amplitudes have been chosen so that $\lambda = 1$ gives $R_x(\pi)$, and $\lambda = 0$ must give $\theta_x = 0$ as the driving fields are off, so varying $\lambda \in (0, 1)$ necessarily gives intermediate values of $\theta_x$. All other parameters are set to the values quoted at the beginning of Sec.~\ref{stateSection}.  Plots of $\DE$ and the AC fields are given in Fig.~\ref{XPulsePlots}.

\begin{figure}
\includegraphics[width=\linewidth]{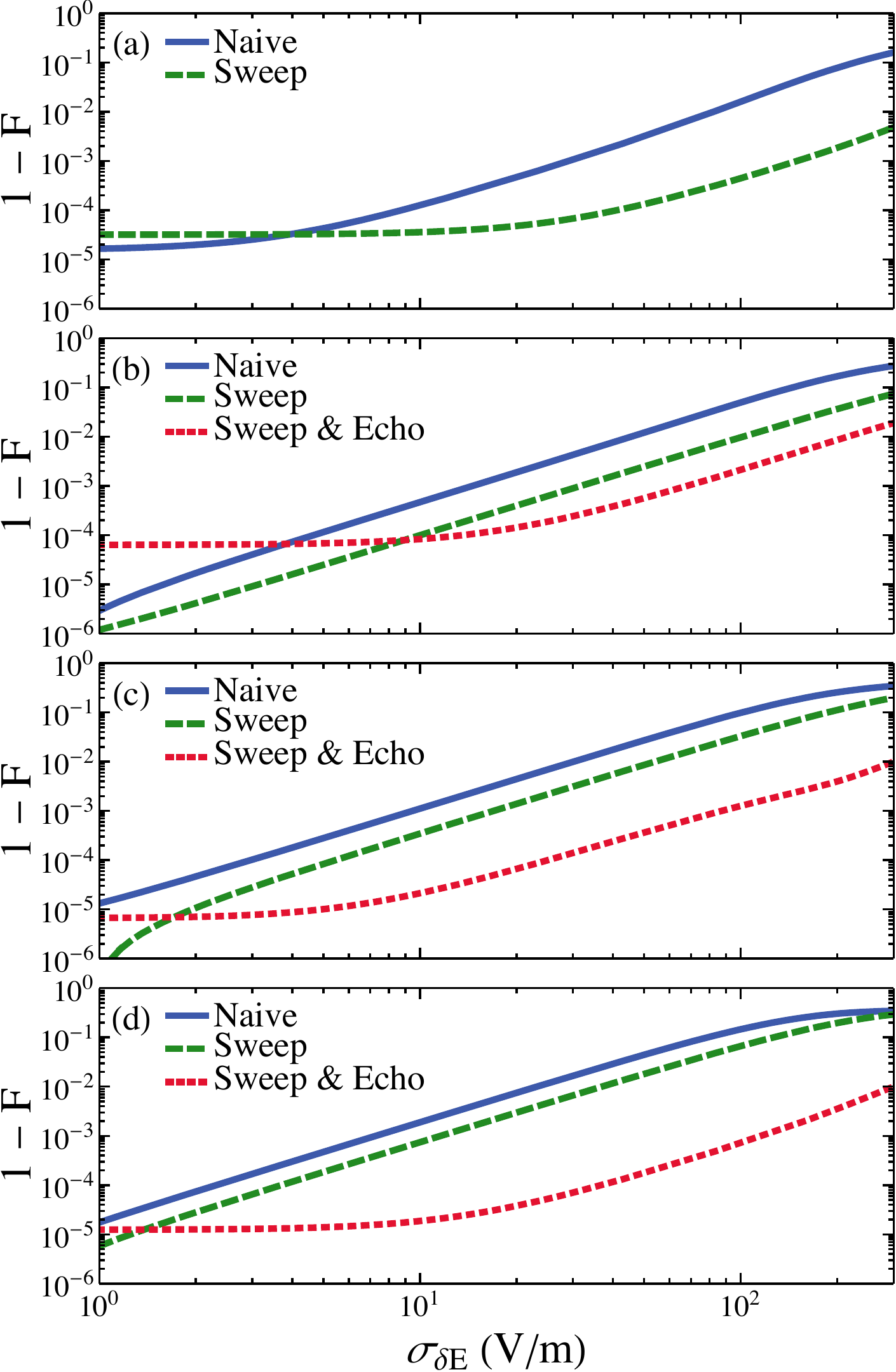}
\caption{Infidelity of simulated $R_x(\theta)$ gates for (a) $\theta = \pi$, (b) $\theta = 3\pi/4$, (c) $\theta = \pi/2$, (d) $\theta = \pi/4$.  The infidelity of the different control schemes described in the main text is plotted for each angle against the charge noise strength included in the simulation.  We show only naive and sweep gates for $\theta = \pi$ because the sweep $R_x(\pi)$ gate is designed so that the errors due to charge noise cancel without the need for echoes.} \label{RxFidelityPlot}
\end{figure}

Figure \ref{RxFidelityPlot}(a) shows the effect of quasi-static charge noise on both a naive $X$-gate as described earlier and on our noise-resistant $X$-gate. Both gates take roughly $\SI{140}{\nano\second}$, including the corrective $Z$-rotations at the beginning and end of the gate.  The noise-resistant gate shows substantially better performance, with an error well below $10^{-3}$ for a noise with r.m.s. of $\SI{100}{\volt\per\meter}$. 

As our protocol uses adiabaticity and involves a sweep, it might seem suggestive of adiabatic passage protocols in NMR ~\cite{Garwood2001}, but it is in fact quite different: here, the sweep does not cause the transition, serving only to suppress the effects of noise.  The gate is also equally adiabatic without it.  The reason that even the naive $X$-gate scheme could have such high fidelity without noise correction is that the large detunings and slowly-changing pulse envelopes ensure that the evolution is very nearly adiabatic.  Using our corrected rotating-frame Hamiltonian $H'$ (see Appendix \ref{AFloquet}), we can find the eigenstates in the middle of a gate's evolution and examine the purity of the evolution operator in the computational subspace to quantify the adiabaticity.  Doing so, we find that the $R_x(\theta)$ gates described above have leakage due to nonadiabaticity near or below $10^{-4}$ at any given point throughout the evolution. However, one complication with our scheme is the fact that if $\varepsilon_0 \approx 2 \omega_E$ at some point while the oscillating electric field is on, there is a weak, sharp two-photon resonance that excites the $|g\rangle$ states to $|e\rangle$ states.  The result of this is that, with our scheme, if $\DE$ is swept over too broad a range, there will be nonadiabaticity of order $10^{-2}$ and a discrepancy between the exact and approximate evolution.  With our parameters, this occurs near $\DE = \pm \SI{2500}{\volt\per\meter}$, so this limits the sweeping range of $\DE$.

We now consider the problem of a general, noise-resistant $R_x(\theta)$ gate. The challenge is to take our noise-vulnerable sweep gate, which gives an evolution of the form $R_z(\theta_{z1} + \theta_{z1}' \dE)R_x(\theta_x)R_z(\theta_{z2} + \theta_{z2}' \dE)$, and find a way to cancel the first-order error terms $\theta_{zi}' \dE$. Once the errors are cancelled, the residual $\theta_{zi}$ terms can be removed by $R_z(\theta)$ gates as before, leaving only the $R_x(\theta_x)$.  For a given $\theta_x$, we can run simulations with different values of $\dE$ and decompose the resulting gates to numerically find $\{\theta_{z1}, \theta'_{z1}, \theta_{z2}, \theta'_{z2}\}$. To apply this protocol in specific experiments, we can use numerical simulations to approximate these parameters, with further optimization using the physical system to account for error.  We find that the first-order approximation for the noise-dependence of the $R_z(\theta)$ components of the decomposition is quite accurate (i.e. there is no need to include a $\theta''_{z1} \dE^2 / 2$ term, for example), indicating that correcting noise to first-order should give a good noise-resistant gate.

The key ingredient in our approach will be a modified, noise-vulnerable $R_z(\theta)$ gate, whose dependence on charge noise we use to cancel the noise in the sweep gate.  If, in an $R_z(\theta)$ gate, we idle at a nominal value of $\DE = 0$ instead of $\DE \ll 0$ (a noise-vulnerable point instead of a noise-resistant point according to Fig.~\ref{dephasingPlot}), we get a rotation error $R_z\Big( \frac{A d e t}{4 \hbar V_t} \dE \Big)$ in addition to the normal $Z$ rotation.  For idling times on the order of tens of $\SI{}{\nano\second}$, this produces an error that is comparable to the dephasing errors $\theta_{zi}'$. If these dephasing angles had the opposite sign as the $\theta_{zi}'$, the solution would be simple: one could apply a gate of this form before and after the sweep gate with the dephasing angles equal to $\theta_{zi}'$ to cancel the noise.  However, the two types of error have the same sign.  Our solution is to use the noise-resistant $R_x(\pi)$ we constructed earlier to flip the sign of this dephasing to use it to cancel the dephasing in the sweep gate.  Here, we are again using $R_z(\theta_z)R_x(\pi) = R_x(\pi)R_z(-\theta_z)$.

The full noise-resistant gate is generated as follows:
\begin{equation}\label{fullRxEquation}
\begin{aligned}
    R_x(\theta_x) =& \overbrace{R_z(\theta_{z1} - \nu_1)}^\text{corrective $R_z$}  \overbrace{R_z(\nu_1 + \theta_{z1}' \dE)}^\text{echo $R_z$} X \\
   & R_z(\theta_{z1} + \theta_{z1}' \dE)R_x(\theta_x)R_z(\theta_{z2} + \theta_{z2}' \dE) \\
    &X \underbrace{R_z(\nu_2 + \theta_{z2}' \dE)}_\text{echo $R_z$} \underbrace{R_z(\theta_{z2} - \nu_2)}_\text{corrective $R_z$}.
\end{aligned}    
\end{equation}
In creating this gate, first one chooses $\theta_x$, which fixes $\lambda$ and the second line of Eq.~\ref{fullRxEquation}, which is a sweep gate. Next, the echo $R_z$ gates are adjusted to cancel the noise terms. All remaining phases are then cancelled by the corrective $R_z$ gates.  In practice, one can avoid the corrected $R_z$ gates contained in the $X$-gates and simply absorb them into the main corrective $R_z$ gates.  The majority of the gate duration comes from the three sweeping gates, each contributing $\SI{120}{\nano\second}$ to the total gate time of $\sim \SI{450}{\nano\second}$.  The noise resistance of $R_x$-gates of various angles is plotted in Fig.~\ref{RxFidelityPlot}.  For charge noise with a r.m.s. of $\SI{100}{\volt\per\meter}$, the full sweep \& echo gates all have error near $10^-3$, well over an order of magnitude better than the naive, non-sweeping gates.

Fig.~\ref{RxFidelityPlot} shows some curves leveling out to somewhat higher infidelity at zero noise. The reason is that the sweep gate was optimized for the extremal value of $\theta = \pi$, with all other angles reached by interpolation. Our optimization only reached $\theta \approx 3.13$ (slightly higher for the naive gate than the sweep gate), so the $\theta = \pi$ gates have higher infidelity at zero noise.  The sweep \& echo gates use $R_x(\pi)$ sweep gates in their pulse sequences, so they also have higher infidelity at zero noise. At realistic noise levels, this small infidelity is negligible.

\section{Two-Qubit Gates} \label{2Qsection}

The spin-charge hybridization obtained by the displacement of the electron from the donor towards the interface induces an electric dipole that can be used for long-range coupling between qubits via a dipole-dipole interaction. As with single-qubit gates, however, it is not obvious how to implement such a gate without leakage into the large two-qubit leakage space.  Ref. \onlinecite{Tosi2017} presents a method that uses an AC magnetic field to couple the nuclear spin qubit to the charge qubit, leading to an i\textsc{swap} gate.  Here we show an alternative method, implementing a \textsc{cphase} gate between two qubits with only an AC electric field.  

First, we must derive the dipole-dipole interaction term.  The electric dipole operator of a qubit depends only on the electron orbital.  We can write the dipole operator of qubit $k$ as
\begin{equation}
\mathbf{p_k} = p_i \ket{i}\bra{i} + p_d \ket{d}\bra{d},
\end{equation}
where $p_i$ and $p_d$ are the effective dipoles when the electron is in the $\ket{i}$ and $\ket{d}$ states, respectively.  Because the electron is at the interface in the $\ket{i}$ state and generally near the donor nucleus in the $\ket{d}$ state, we expect $p_i \approx de$ and $p_d \approx 0$. We use these approximations for the remainder of this section for simplicity.  The interaction energy of two qubits, 1 and 2, separated by a displacement $\mathbf{r}$ is given by
\begin{equation}
V_{dip} = \frac{\mathbf{p_1}\cdot\mathbf{p_2} - 3(\mathbf{p_1}\cdot\mathbf{r})(\mathbf{p_2}\cdot\mathbf{r})/r^2}{4 \pi \epsilon_0 \epsilon_r r^3},
\end{equation}
where $\epsilon_0$ is the vacuum permittivity and $\epsilon_r$ is the dielectric constant of the material ($\epsilon_r = 11.7$ for silicon).  We assume that all qubits will be fabricated with their dipoles perpendicular to the surface on which they are arrayed.  In this case $\mathbf{p_k}\cdot\mathbf{r} = 0$, so the interaction simplifies to
\begin{equation}
V_{dip} = \frac{e^2 d^2 \ket{i_1 i_2}\bra{i_1 i_2}}{4 \pi \epsilon_0 \epsilon_r r^3}.
\end{equation}

We can then model the system with the Hamiltonian $H_{2q} = H \otimes \mathds{1} + \mathds{1} \otimes H + V_{dip}$.  The lowest four eigenstates will be very similar to the tensor product of the individual qubit eigenstates without $V_{dip}$, so we call them $\{\upup, \updown, \downup, \downdown\}$ and use them as the computational states.  Our goal is to use the dipole-dipole interaction to implement a \textsc{cphase} gate between two adjacent qubits.  We use a method similar to our $R_z(\theta)$ scheme: we use electric fields to cause an adiabatic evolution in which the four computational states shift in energy and accumulate phases without transitions between eigenstates.  The evolution operator $U$ at the end of the evolution will be of the form
\begin{equation}
\begin{aligned}
U =& e^{i \alpha} \upup \upupBra + e^{i \beta} \updown \updownBra\\
    &+ e^{i \gamma} \downup \downupBra + e^{i \delta} \downdown \downdownBra.
\end{aligned}    
\end{equation}
If we apply the local rotations $R_{z}(\gamma - \alpha)$ to qubit 1 and $R_{z}(\beta - \alpha)$ to qubit 2, and ignore a global phase of $(\beta + \gamma)/2$, we get
\begin{equation}
\begin{aligned}
   U' =& \upup \upupBra + \updown \updownBra  \\
    &+ \downup \downupBra + e^{i (\alpha - \beta - \gamma + \delta)} \downdown \downdownBra.
\end{aligned}    
\end{equation}
Thus if $\phi \equiv \alpha - \beta - \gamma + \delta$ is nonzero, we obtain a \textsc{cphase} gate with angle $\phi$.

Assuming that the evolution will be adiabatic, we can find the angles $\alpha, \beta, \gamma, \delta$ and thus calculate $\phi$ using an expression analogous to Eq.~\ref{phiZEquation}, where we integrate each state's energy over the course of the evolution.  Using our approximation, the energies are given by
\begin{equation} \label{EabEquation}
    E_{ab} \approx E_a + E_b + \frac{e^2 d^2}{4 \pi \epsilon_0 \epsilon_r r^3} \cdot \lvert\braket{i}{a}\rvert^2 \cdot \lvert\braket{i}{b}\rvert^2,
\end{equation}
where $a \in \{\Tilde{\Uparrow}_1, \Tilde{\Downarrow}_1\}$ and $b \in \{\Tilde{\Uparrow}_2, \Tilde{\Downarrow}_2\}$.  When one integrates the energies given by Eq.~\ref{EabEquation} to find $\alpha, \beta, \gamma, \delta$ and get an expression for $\phi$, the single-qubit energies ($E_a$ and $E_b$ in Eq.~\ref{EabEquation}) cancel, and the remaining terms can be simplified to
\begin{equation} \label{CZphaseIntegral}
\begin{aligned}
    \phi \approx& \int^T_0 dt \frac{-e^2 d^2}{4 \pi \epsilon_0 \epsilon_r r^3}\cdot (\lvert\bra{i}\Tilde{\Uparrow}_1\rangle|^2 -  \lvert\bra{i}\Tilde{\Downarrow}_1\rangle|^2)\\
    &\times (\lvert\bra{i}\Tilde{\Uparrow}_2\rangle|^2 -  \lvert\bra{i}\Tilde{\Downarrow}_2\rangle|^2).
\end{aligned}    
\end{equation}

A key first question is how the dipole-dipole interaction affects the qubits while idling.  For our choice of idling states, both qubit states have the electron in the $\ket{g}$ state, so $\lvert\bra{i}\Tilde{\Uparrow}\rangle|^2 = \lvert\bra{i}\Tilde{\Downarrow}\rangle|^2$.  This implies that $\phi \approx 0$.  Furthermore, even if only one qubit is idling, the corresponding factor in Eq.~\ref{CZphaseIntegral} will be zero, so there will still be no accumulated $\phi$.  The fact that there is no entanglement when at least one of the qubits is idling is a major advantage of this choice of idling point and basis states.

Eq.~\ref{CZphaseIntegral} implies that both qubits' two computational states must have different average dipoles in order for a nonzero $\phi$ to accumulate.  We can achieve this with only oscillating electric fields at each qubit.  The key ingredient is the fact that the two computational states have different ground-excited electron orbital splittings due to the hyperfine interaction: $E_{\edu} - E_{\gdu} \approx \varepsilon_0 - A/4 + \langle A \rangle/2$, and $E_{\edd} - E_{\gdd} \approx \varepsilon_0 + A/4 - \langle A \rangle/2$.  Driving $E_{ac}$ for both qubits near, for example, the $\down$ state's electron orbital transition energy but far detuned from the $\up$ state's will give the $\down$ state a significant $|e\rangle$ component, creating a difference in dipole and leading to a nonzero $\phi$ according to Eq.~\ref{CZphaseIntegral}.

To give an example of a square-pulse implementation of this idea, we assume a qubit spacing of $r = \SI{500}{\nano\meter}$, and setting $E_{a} = \SI{30}{\volt\per\meter}$, $\omega_E = \varepsilon_0 + A/4 - \langle A \rangle/2 + 2\pi \cdot \SI{5}{\mega\hertz}$, $B_{ac} = 0$,  $\DE = \SI{2000}{\volt\per\meter}$, while keeping the other parameters the same as before, we find from $H'$ (see Appendix~\ref{AFloquet}) that $\up \approx 0.923 \gdu - 0.368 \edu$ and $\down \approx 0.711 \gdd - 0.703 \edd$; the $\down$ state is driven closer to its electron orbital resonance and thus has a greater $\ket{e}$ component.  We have changed $\DE$ from idling because in Eq.~\ref{RWAHamiltonianEquation}, the oscillating electric field is attenuated by a coefficient of $V_t / \varepsilon_0$, so decreasing $|\DE|$ will allow a smaller driving electric field to achieve the same effect.  We can now use Eq.~\ref{CZphaseIntegral} and the definitions of the $\{\ket{g}, \ket{e}\}$ states to find that $\phi/T \approx 2\pi \cdot \SI{1.9}{MHz}$.  This yields a \textsc{cz} gate, with $\phi = \pi$, in $\sim \SI{500}{\nano\second}$.

Performing a similar gate with realistic, smooth pulses is slightly more complicated due to the need to vary the AC electric field adiabatically and to the fact that the ground-excited splitting is changed slightly by the presence of a second qubit, requiring an adjustment in $\omega_E$.  Like the $R_z$ gates, we parameterize \textsc{cphase} gates in terms of their total duration $T$ and calculate the resulting angle $\phi$.  We choose $\DE(t) = \DE_{idle} + l(t, \tau_1, -\DE_{idle} + D, \tau_1 + \tau_{ac}, -\DE_{idle} + D, T)$, $E_{a}(t) = E_{max} \cdot w(t - \tau_1, \tau_2, \tau_{ac})$, and $\omega_E = \varepsilon_0 + A/4 - \langle A \rangle/2 - 2\pi \cdot \SI{10}{\mega\hertz}$, where $\DE_{idle} = \SI{10000}{\volt\per\meter}$, $D = \SI{2000}{\volt\per\meter}$ is the value $\DE$ moves to during the gate, $E_{max} = (\SI{40}{\volt\per\meter}) \cdot \min(1, (T/\SI{300}{\nano\second})^2)$ is the max amplitude of $E_{ac}$, $\tau_1 = \SI{5}{\nano\second}$ is the setup time, $\tau_{ac} = T - 2 \tau_1$ is the time when $\DE$ is constant and $E_{a}$ is nonzero, and $\tau_2 = \min(\SI{300}{\nano\second}, \tau_{ac}/2)$ is the ramp-up time of $E_{a}$.  When $T$ is small, $E_{max}$ is smaller to maintain adiabaticity, and the ramp-up time $\tau_2$ must also be small to fit within $T$.  These pulses are very similar in shape to those used for the sweep $R_x$ gate, except that here $B_{ac} = 0$ and $\DE$ is constant for most of the duration of the gate.

\begin{figure}
\includegraphics[width=\linewidth]{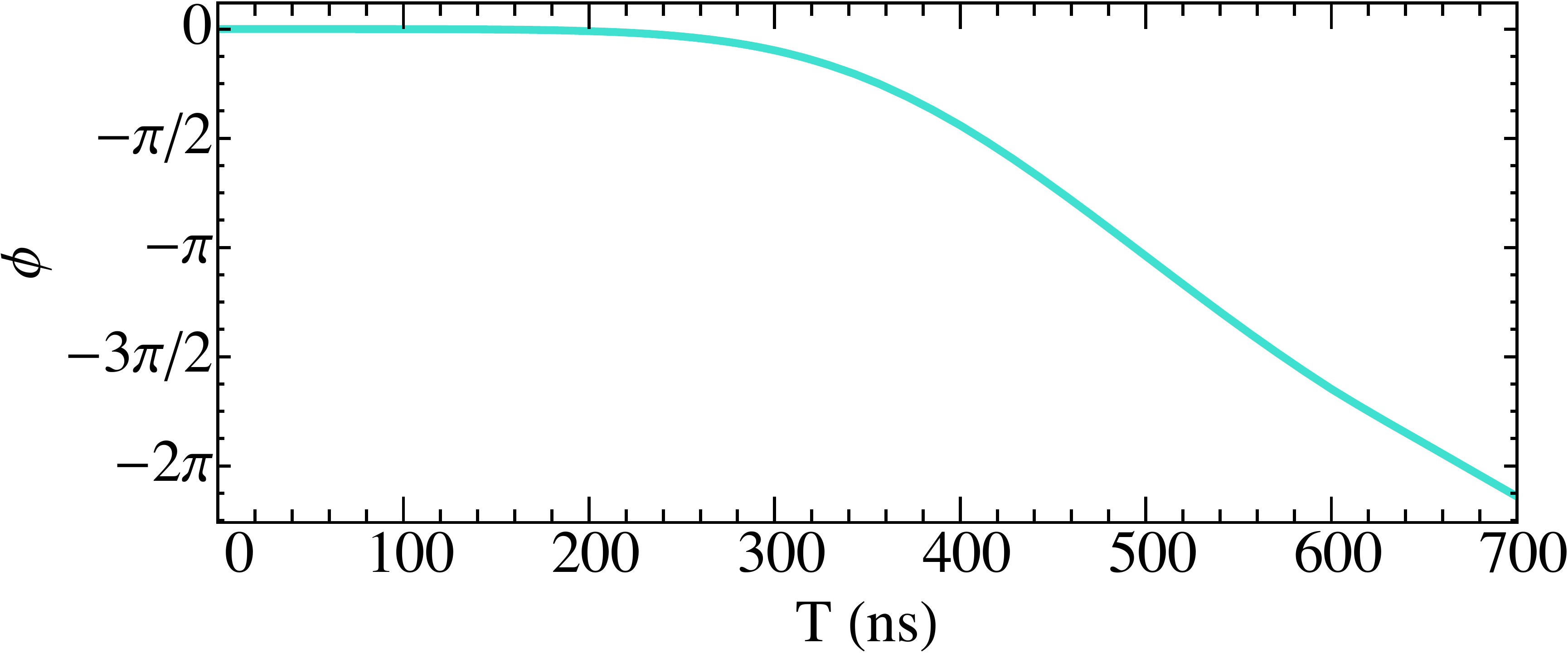}
\caption{The phase of a \textsc{cphase} gate after subtracting local phases as a function of gate duration.  The infidelity due to nonadiabaticity is always below $10^{-3}$ and could be decreased further by attenuating the driving field and increasing gate time.  A \textsc{cz} gate, with $\phi = \pi$, is implemented when $T = \SI{494}{\nano\second}$.} \label{CZphasePlot}
\end{figure}

Figure \ref{CZphasePlot} shows the value of $\phi$ of our \textsc{cphase} gate as the gate duration varies.  Including the local $Z$-rotations to correct the local phases, a \textsc{cz} gate takes $\sim \SI{500}{\nano\second}$ and an arbitrary \textsc{cphase} gate takes less than $\SI{750}{\nano\second}$, fast enough that hundreds or thousands of two-qubit gates can be implemented within the decoherence time.

To be truly practical, this \textsc{cphase} gate scheme must have a degree of charge-noise-resistance.  This is a significant challenge because of the requirement that the two states of each qubit have different but precisely known dipole moments.  Charge noise $\dE$ will lead to a perturbation of the form $-(d e \dE / 2 \hbar) \mathbf{\tau}^{id}_{z}$, which, taken to first order, will then perturb the energies of the two qubit states by different amounts. This change in the qubit energy splitting causes significant dephasing even for small $\dE$. Finding a way to make this entangling gate scheme or an alternative scheme noise-resistant remains an open problem and will be the subject of future work.

\section{Conclusion} \label{conclusion}
To conclude, we have introduced quantum control schemes to implement fast high-fidelity single- and two-qubit gates for $^{31}$P  nuclear spin qubits in silicon. We presented protocols for implementing arbitrary $R_z(\theta)$ and $R_x(\theta)$ single-qubit gates, which can be combined to make an arbitrary, noise-resistant single-qubit gate in under $\SI{500}{\nano\second}$. For typical charge noise levels of $\SI{100}{\volt\per\meter}$, our procedure achieves fidelities over $99.99\%$ for arbitrary $Z$-rotations and fidelities over $99.9\%$ for arbitrary $X$-rotations. This is well above the threshold error rate of some quantum error correction codes, e.g. the surface code~\cite{Fowler2012a,Raussendorf2007}. We choose a computational basis such that two qubits are only entangled when both are driven simultaneously, allowing for single-qubit gates to be performed without crosstalk from adjacent idling qubits. We also introduced a method for implementing two-qubit controlled-phase gates with arbitrary phases that take less than $\SI{750}{\nano\second}$ for an inter-qubit distance of $\SI{0.5}{\micro\meter}$, using only an oscillating electric field. These results are immediately relevant to ongoing experiments on donor-based nuclear spin qubits.

\section*{Acknowledgments}

This work is supported by the Army Research Office (W911NF-17-0287).

\appendix
\section{Multi-Frequency Floquet Theory} \label{AFloquet}

This appendix shows how to use multi-frequency Floquet theory and second-order perturbation theory to find $H'$, an accurate time-independent approximation to the system Hamiltonian.  The first step is to find $\{\Tilde{H}_{\omega_j}\}_j$, the frequency components of $\Tilde{H}$ such that $\Tilde{H} = \sum_j \Tilde{H}_{\omega_j} e^{i \omega_j t}$.  The frequencies present are $\{\omega_i\}_i$ = $\{0, \pm\omega_E, \pm 2 \omega_E, \pm 2 \omega_B, \pm (2 \omega_B - \omega_E)\}$.  $\Tilde{H}_0$ is given in Section \ref{derivationSection}, and the rest are given by the following expressions, noting that $\Tilde{H}_j = \Tilde{H}^\dagger_{-j}$:
\begin{equation}
\begin{aligned}
    \Tilde{H}_{\omega_E} &= \frac{A}{4}(\Sx - i\Sy)(\Ix + i\Iy) - \frac{B_{a}(t) \gamma_n}{4} (\Ix + i\Iy) \\
   & + \Big(\frac{\langle A \rangle}{2} - \frac{A}{4}\Big)(\Sx - i\Sy)(\Ix + i\Iy)\\
   &- \frac{A V_t}{4 \varepsilon_0}(\sigx - i\sigy)\Sz\Iz + \frac{V_t B_0 \gamma_e \Delta\gamma}{4 \varepsilon_0}(\sigx - i\sigy)\Sz\\
   &- \frac{d^2 e^2 \DE E_{a}(t)}{4 \hbar^2 \varepsilon_0} \sigz.\\
    \Tilde{H}_{2\omega_E} &= -\frac{A V_t}{8 \varepsilon_0}(\sigx - i\sigy)(\Sx - i\Sy)(\Ix + i\Iy) \\
    &- \frac{V_t d e E_{a}(t)}{8 \hbar \varepsilon_0}(\sigx - i\sigy),\\
    \Tilde{H}_{2\omega_B} &= \frac{B_{a}(t) \gamma_e}{4}(\Sx - i\Sy),\\
    \Tilde{H}_{2\omega_B - \omega_E} &= -\frac{B_{a}(t) \gamma_n}{4}(\Ix - i\Iy).
\end{aligned}    
\end{equation}

A system with one driving frequency can be analyzed in a time-independent way by considering a Floquet Hamiltonian \cite{floquetTheory}, an infinite Hamiltonian whose basis is the tensor product of the original basis with the space of integers, with each integer representing a Fourier component of the solution.  When the RWA fails because the driving is strong, perturbation theory on the Floquet Hamiltonian can derive corrections (like in the Bloch-Siegert shift) that preserve the time-independent approximation.  We do a similar process here, using the multi-frequency Floquet formalism given in Ref. \onlinecite{multiFreq}.

First, we construct the multi-frequency Floquet Hamiltonian $H_F$.  Because the base Hamiltonian is 8$\times$8 and there are 9 distinct frequencies, the Floquet Hamiltonian will be 72$\times$72.  The diagonal will be populated with copies of $\Tilde{H}_0$ shifted by unique frequencies - just the frequencies present in $\Tilde{H}$ in our truncated approximation - and the off-diagonal will have the components of $\Tilde{H}$ whose frequencies are the differences of the shifts of the matrices along the diagonal.  $H_F$, truncated to one order in each frequency, is shown in Eq.~\ref{fullHF}, in which each matrix element represents the corresponding $8\times8$ matrix.  It will turn out that only the matrices that are part of the diagonal or the central row or column will matter to our approximation, but we include them all for completeness.
\begin{widetext}
\setcounter{MaxMatrixCols}{20}
\begin{equation}\label{fullHF}
H_F = \begin{bmatrix}
 \Tilde{H}_0 - 2\omega_E & 0 & \Tilde{H}_{-\omega_E} & 0 & H_{-2\omega_E} & 0 & 0 & 0 & 0 \\
 0 & \Tilde{H}_0 - 2\omega_B & \Tilde{H}_{-2\omega_B + \omega_E} & \Tilde{H}_{-\omega_E} & \Tilde{H}_{-2 \omega_E} & 0 & 0 & 0 & 0 \\
\Tilde{H}_{\omega_E} & \Tilde{H}_{2\omega_B - \omega_E} & \Tilde{H}_0 - \omega_E & 0 & \Tilde{H}_{-\omega_E} & \Tilde{H}_{-2\omega_B} & \Tilde{H}_{-2\omega_B} & 0 & 0 \\
 0 & H_{\omega_E} & 0 & \Tilde{H}_0 - 2\omega_B + \omega_E & \Tilde{H}_{-2\omega_E + \omega_E} & 0 & \Tilde{H}_{-2\omega_B} & 0 & 0 \\
 \Tilde{H}_{2\omega_E} & \Tilde{H}_{2\omega_B} & \Tilde{H}_{\omega_E} & \Tilde{H}_{2\omega_B - \omega_E} & \Tilde{H}_0 & \Tilde{H}_{-2\omega_B + \omega_E} & \Tilde{H}_{\omega_E} & \Tilde{H}_{2\omega_B} & \Tilde{H}_{2\omega_E} \\
 0 & 0 & \Tilde{H}_{2\omega_B} & 0 & \Tilde{H}_{2\omega_B - \omega_E} & \Tilde{H}_0 + 2\omega_B - \omega_E & 0 & \Tilde{H}_{\omega_E} & 0 \\
 0 & 0 & \Tilde{H}_{2\omega_E} & \Tilde{H}_{2\omega_B} & \Tilde{H}_{\omega_E} & 0 & \Tilde{H}_0 + \omega_E & \Tilde{H}_{-2\omega_B + \omega_E} & \Tilde{H}_{-\omega_E} \\
 0 & 0 & 0 & 0 & \Tilde{H}_{2\omega_B} & \Tilde{H}_{\omega_E} & \Tilde{H}_{2\omega_B - \omega_E} & \Tilde{H} + 2 \omega_B & 0 \\
 0 & 0 & 0 & 0 & \Tilde{H}_{2\omega_E} & 0 & \Tilde{H}_{\omega_E} & 0 & \Tilde{H}_0 + 2\omega_E,
\end{bmatrix}
\end{equation}
\end{widetext}

If we note that the eigenvalues of $\Tilde{H}_0$ will be much smaller than either $\omega_E$ or $\omega_B$ (by a factor of over $\sim 10$ with our parameters), the dynamics can be easily approximated.  The matrices along the diagonal are all very well-separated in energy from $\Tilde{H}_0$, so we can treat all the off-diagonal elements in $H_F$ as a perturbation and derive an effective $H_0$ using second-order quasi-degenerate perturbation theory, also called a Schrieffer-Wolff transformation \cite{schriefferWolff}.

We choose the 8-dimensional subspace through the unshifted $\Tilde{H}_0$ at the center of $H_F$ as the target subspace of the Schrieffer-Wolff transformation.  We now define $H^{(0)}_F$ as the diagonal (the diagonal, not just a block diagonal) of $H_F$, and define the perturbation $H'_F$ so $H_F = H^{(0)}_F + H'_F$.  Our goal is to apply a small transformation such that the target subspace becomes decoupled from the rest of the Floquet Hamiltonian, leaving an effective Hamiltonian $H_{eff}$.  As derived in Ref. \onlinecite{schriefferWolff}, this is given by
\begin{equation}
    H_{eff} = H^{(0)} + H^{(1)} + H^{(2)} + \ldots ,
\end{equation}
where
\begin{equation}
\begin{aligned}
    H^{(0)}_{m m'} &= \tilde{H_F^0}_{m m'},\\
    H^{(1)}_{m m'} &= {H'_F}_{m m'},\\
   H^{(2)}_{m m'} &= \frac{1}{2} \sum_l {H'_F}_{m l} {H'_F}_{l m'} \Bigg[ \frac{1}{E_m - E_l} + \frac{1}{E_{m'} - E_l} \Bigg],
\end{aligned}
\end{equation}
where the states $m$ and $m'$ are states within the target subspace and $E_l$ is the energy of state $l$ before the perturbation, which is simply ${H^{0}_F}_{ll}$ because $H^{0}_F$ is diagonal.  Note that $H^{(0)}$ is just the diagonal of $\Tilde{H}_0$ and $H^{(1)}$ is the off-diagonal, so $H^{(0)} + H^{(1)} = \Tilde{H}_0$.  Summing the 0th-, 1st- and 2nd-order terms in $H_{eff}$ gives $H'$, the accurate, time-independent approximation mentioned in Section \ref{derivationSection}, with typical fidelity $>.9999$ to the exact evolution for our gates.

\section{Qubit Dephasing Rates} \label{Adephasing}
We here derive the dephasing rate of the qubit in our scheme due to charge noise.  The critical value is the derivative of the qubit energy difference with respect to the electric field,
\begin{equation}
    \frac{d \delta_q}{d \DE} = -\frac{A d e {V_t}^2}{4 \hbar {\varepsilon_0}^3}.
\end{equation}
When $d e \DE / \hbar \gg V_t$, $\varepsilon_0 \approx d e \DE / \hbar$, so the above equation simplifies to
\begin{equation}
    \frac{d \delta_q}{d \DE} \approx - \frac{A \hbar^2 {V_t}^2}{4 d^2 e^2 \DE^3}.
\end{equation}
With our idling parameters, this is $2\pi \cdot 70 \text{ Hz/Vm}^{-1}$, so assuming a typical noise in $\DE$ of $\SI{100}{\volt\per\meter}$ gives a dephasing time on the order of $\SI{.1}{\milli\second}$.

\section{Effect of a Time-Dependent $\{\ket{g}, \ket{e}\}$ Basis} \label{Age}
When we transform the system Hamiltonian from the ($|i\rangle$, $|d\rangle$) basis to the ($|g\rangle$, $|e\rangle$) basis in Section \ref{systemSection}, we treat $\DE$ as static and don't include a $ - i \Lambda \dot{\Lambda}^\dagger$ term for that change of basis, even though $\DE$ is not constant during gates.  Here we give that correction and show that it is negligible, a conclusion which we checked by comparing our simulations to simulations in the original ($|i\rangle$, $|d\rangle$) basis.

The transformation from the ($|i\rangle$, $|d\rangle$) basis to the ($|g\rangle$, $|e\rangle$) is given by
\begin{equation}
    \Lambda = \frac{1}{\sqrt{2}} \sqrt{1 + \frac{d e \DE}{\hbar \varepsilon_0}}\mathds{1} - \frac{i}{\sqrt{2}}\sqrt{1 - \frac{d e \DE}{\hbar \varepsilon_0}} \sigy,
\end{equation}
yielding
\begin{equation} \label{missingTermEquation}
    -i \Lambda \dot{\Lambda}^\dagger = \frac{d e V_t}{2 \hbar \varepsilon_0^2} \sigy \cdot \frac{d \DE}{dt}.
\end{equation}
The factor $d e V_t / 2 \hbar \varepsilon^2_0$ reaches a maximum of $\sim \SI{2e-3}{\meter\per\volt}$ when $\DE = 0$.  

There are two situations where this extra term could be problematic.  First, during the very rapid shifts in electric field at the start and end of the $Z$- and $X$-rotations, the large term could cause unwanted $|g\rangle - |e\rangle$ coupling.  Second, in the middle of an $X$-rotation, the system dynamics are fairly sensitive to the detunings between states, so a smaller added term from the more-slowly changing electric field could be problematic.

For the first case, at maximum, $|d\DE/dt| \sim \SI{e13}{\volt\per\meter\per\second}$, so Eq.~ \ref{missingTermEquation} is of order $2 \pi \cdot \SI{1}{\giga\hertz}$, while $\varepsilon_0 \sim 2 \pi \cdot \SI{5}{\giga\hertz}$.  This term is big enough to be problematic if it were turned on suddenly, but the fact that it waxes and wanes over the course of several $\SI{}{\nano\second}$ preserves adiabaticity, and it is only large for a very short time, so it does not significantly affect the gate.

In the second case, in the middle of an X-rotation, the term is of order $2 \pi \cdot \SI{5}{\mega\hertz}$.  This is two orders of magnitude smaller than the detunings and much smaller than the energy splitting $\varepsilon_0$ of the states it couples.  If it were included in the derivation of $H'$, it would only contribute as a small correction.  We have confirmed that it does not significantly affect the results of our simulations.

\bibliography{refs} 

\begin{thebibliography}{34}%
\makeatletter
\providecommand \@ifxundefined [1]{%
 \@ifx{#1\undefined}
}%
\providecommand \@ifnum [1]{%
 \ifnum #1\expandafter \@firstoftwo
 \else \expandafter \@secondoftwo
 \fi
}%
\providecommand \@ifx [1]{%
 \ifx #1\expandafter \@firstoftwo
 \else \expandafter \@secondoftwo
 \fi
}%
\providecommand \natexlab [1]{#1}%
\providecommand \enquote  [1]{``#1''}%
\providecommand \bibnamefont  [1]{#1}%
\providecommand \bibfnamefont [1]{#1}%
\providecommand \citenamefont [1]{#1}%
\providecommand \href@noop [0]{\@secondoftwo}%
\providecommand \href [0]{\begingroup \@sanitize@url \@href}%
\providecommand \@href[1]{\@@startlink{#1}\@@href}%
\providecommand \@@href[1]{\endgroup#1\@@endlink}%
\providecommand \@sanitize@url [0]{\catcode `\\12\catcode `\$12\catcode
  `\&12\catcode `\#12\catcode `\^12\catcode `\_12\catcode `\%12\relax}%
\providecommand \@@startlink[1]{}%
\providecommand \@@endlink[0]{}%
\providecommand \url  [0]{\begingroup\@sanitize@url \@url }%
\providecommand \@url [1]{\endgroup\@href {#1}{\urlprefix }}%
\providecommand \urlprefix  [0]{URL }%
\providecommand \Eprint [0]{\href }%
\providecommand \doibase [0]{http://dx.doi.org/}%
\providecommand \selectlanguage [0]{\@gobble}%
\providecommand \bibinfo  [0]{\@secondoftwo}%
\providecommand \bibfield  [0]{\@secondoftwo}%
\providecommand \translation [1]{[#1]}%
\providecommand \BibitemOpen [0]{}%
\providecommand \bibitemStop [0]{}%
\providecommand \bibitemNoStop [0]{.\EOS\space}%
\providecommand \EOS [0]{\spacefactor3000\relax}%
\providecommand \BibitemShut  [1]{\csname bibitem#1\endcsname}%
\let\auto@bib@innerbib\@empty
\bibitem [{\citenamefont {Saeedi}\ \emph {et~al.}(2013)\citenamefont {Saeedi},
  \citenamefont {Simmons}, \citenamefont {Salvail}, \citenamefont {Dluhy},
  \citenamefont {Riemann}, \citenamefont {Abrosimov}, \citenamefont {Becker},
  \citenamefont {Pohl}, \citenamefont {Morton},\ and\ \citenamefont
  {Thewalt}}]{Saeedi2013}%
  \BibitemOpen
  \bibfield  {author} {\bibinfo {author} {\bibfnamefont {Kamyar}\ \bibnamefont
  {Saeedi}}, \bibinfo {author} {\bibfnamefont {Stephanie}\ \bibnamefont
  {Simmons}}, \bibinfo {author} {\bibfnamefont {Jeff~Z.}\ \bibnamefont
  {Salvail}}, \bibinfo {author} {\bibfnamefont {Phillip}\ \bibnamefont
  {Dluhy}}, \bibinfo {author} {\bibfnamefont {Helge}\ \bibnamefont {Riemann}},
  \bibinfo {author} {\bibfnamefont {Nikolai~V.}\ \bibnamefont {Abrosimov}},
  \bibinfo {author} {\bibfnamefont {Peter}\ \bibnamefont {Becker}}, \bibinfo
  {author} {\bibfnamefont {H.-J.}\ \bibnamefont {Pohl}}, \bibinfo {author}
  {\bibfnamefont {John J.~L.}\ \bibnamefont {Morton}}, \ and\ \bibinfo {author}
  {\bibfnamefont {Mike L.~W.}\ \bibnamefont {Thewalt}},\ }\bibfield  {title}
  {\enquote {\bibinfo {title} {{Room-Temperature Quantum Bit Storage Exceeding
  39 Minutes Using Ionized Donors in Silicon-28}},}\ }\href {\doibase
  10.1126/science.1239584} {\bibfield  {journal} {\bibinfo  {journal}
  {Science}\ }\textbf {\bibinfo {volume} {342}},\ \bibinfo {pages} {830--833}
  (\bibinfo {year} {2013})}\BibitemShut {NoStop}%
\bibitem [{\citenamefont {Muhonen}\ \emph {et~al.}(2014)\citenamefont
  {Muhonen}, \citenamefont {Dehollain}, \citenamefont {Laucht}, \citenamefont
  {Hudson}, \citenamefont {Kalra}, \citenamefont {Sekiguchi}, \citenamefont
  {Itoh}, \citenamefont {Jamieson}, \citenamefont {McCallum}, \citenamefont
  {Dzurak},\ and\ \citenamefont {Morello}}]{Muhonen2014}%
  \BibitemOpen
  \bibfield  {author} {\bibinfo {author} {\bibfnamefont {Juha~T.}\ \bibnamefont
  {Muhonen}}, \bibinfo {author} {\bibfnamefont {Juan~P.}\ \bibnamefont
  {Dehollain}}, \bibinfo {author} {\bibfnamefont {Arne}\ \bibnamefont
  {Laucht}}, \bibinfo {author} {\bibfnamefont {Fay~E.}\ \bibnamefont {Hudson}},
  \bibinfo {author} {\bibfnamefont {Rachpon}\ \bibnamefont {Kalra}}, \bibinfo
  {author} {\bibfnamefont {Takeharu}\ \bibnamefont {Sekiguchi}}, \bibinfo
  {author} {\bibfnamefont {Kohei~M.}\ \bibnamefont {Itoh}}, \bibinfo {author}
  {\bibfnamefont {David~N.}\ \bibnamefont {Jamieson}}, \bibinfo {author}
  {\bibfnamefont {Jeffrey~C.}\ \bibnamefont {McCallum}}, \bibinfo {author}
  {\bibfnamefont {Andrew~S.}\ \bibnamefont {Dzurak}}, \ and\ \bibinfo {author}
  {\bibfnamefont {Andrea}\ \bibnamefont {Morello}},\ }\bibfield  {title}
  {\enquote {\bibinfo {title} {{Storing quantum information for 30 seconds in a
  nanoelectronic device}},}\ }\href {\doibase 10.1038/nnano.2014.211}
  {\bibfield  {journal} {\bibinfo  {journal} {Nat. Nanotechnol.}\ }\textbf
  {\bibinfo {volume} {9}},\ \bibinfo {pages} {986--991} (\bibinfo {year}
  {2014})}\BibitemShut {NoStop}%
\bibitem [{\citenamefont {Vandersypen}\ and\ \citenamefont
  {Chuang}(2005)}]{Vandersypen2004a}%
  \BibitemOpen
  \bibfield  {author} {\bibinfo {author} {\bibfnamefont {L.~M.~K.}\
  \bibnamefont {Vandersypen}}\ and\ \bibinfo {author} {\bibfnamefont {I.~L.}\
  \bibnamefont {Chuang}},\ }\bibfield  {title} {\enquote {\bibinfo {title}
  {{NMR techniques for quantum control and computation}},}\ }\href {\doibase
  10.1103/RevModPhys.76.1037} {\bibfield  {journal} {\bibinfo  {journal} {Rev.
  Mod. Phys.}\ }\textbf {\bibinfo {volume} {76}},\ \bibinfo {pages}
  {1037--1069} (\bibinfo {year} {2005})}\BibitemShut {NoStop}%
\bibitem [{\citenamefont {Jones}(2011)}]{Jones2010}%
  \BibitemOpen
  \bibfield  {author} {\bibinfo {author} {\bibfnamefont {Jonathan~A.}\
  \bibnamefont {Jones}},\ }\bibfield  {title} {\enquote {\bibinfo {title}
  {{Quantum computing with NMR}},}\ }\href {\doibase
  10.1016/j.pnmrs.2010.11.001} {\bibfield  {journal} {\bibinfo  {journal}
  {Prog. Nucl. Magn. Reson. Spectrosc.}\ }\textbf {\bibinfo {volume} {59}},\
  \bibinfo {pages} {91--120} (\bibinfo {year} {2011})}\BibitemShut {NoStop}%
\bibitem [{\citenamefont {Asaad}\ \emph {et~al.}(2019)\citenamefont {Asaad},
  \citenamefont {Mourik}, \citenamefont {Joecker}, \citenamefont {Johnson},
  \citenamefont {Baczewski}, \citenamefont {Firgau}, \citenamefont
  {M{\c{a}}dzik}, \citenamefont {Schmitt}, \citenamefont {Pla}, \citenamefont
  {Hudson}, \citenamefont {Itoh}, \citenamefont {McCallum}, \citenamefont
  {Dzurak}, \citenamefont {Laucht},\ and\ \citenamefont {Morello}}]{Asaad2019}%
  \BibitemOpen
  \bibfield  {author} {\bibinfo {author} {\bibfnamefont {Serwan}\ \bibnamefont
  {Asaad}}, \bibinfo {author} {\bibfnamefont {Vincent}\ \bibnamefont {Mourik}},
  \bibinfo {author} {\bibfnamefont {Benjamin}\ \bibnamefont {Joecker}},
  \bibinfo {author} {\bibfnamefont {Mark A~I}\ \bibnamefont {Johnson}},
  \bibinfo {author} {\bibfnamefont {Andrew~D}\ \bibnamefont {Baczewski}},
  \bibinfo {author} {\bibfnamefont {Hannes~R}\ \bibnamefont {Firgau}}, \bibinfo
  {author} {\bibfnamefont {Mateusz~T}\ \bibnamefont {M{\c{a}}dzik}}, \bibinfo
  {author} {\bibfnamefont {Vivien}\ \bibnamefont {Schmitt}}, \bibinfo {author}
  {\bibfnamefont {Jarryd~J}\ \bibnamefont {Pla}}, \bibinfo {author}
  {\bibfnamefont {Fay~E}\ \bibnamefont {Hudson}}, \bibinfo {author}
  {\bibfnamefont {Kohei~M.}\ \bibnamefont {Itoh}}, \bibinfo {author}
  {\bibfnamefont {Jeffrey~C.}\ \bibnamefont {McCallum}}, \bibinfo {author}
  {\bibfnamefont {Andrew~S}\ \bibnamefont {Dzurak}}, \bibinfo {author}
  {\bibfnamefont {Arne}\ \bibnamefont {Laucht}}, \ and\ \bibinfo {author}
  {\bibfnamefont {Andrea}\ \bibnamefont {Morello}},\ }\bibfield  {title}
  {\enquote {\bibinfo {title} {{Coherent electrical control of a single
  high-spin nucleus in silicon}},}\ }\href {http://arxiv.org/abs/1906.01086} {\
   (\bibinfo {year} {2019})},\ \Eprint {http://arxiv.org/abs/1906.01086}
  {arXiv:1906.01086} \BibitemShut {NoStop}%
\bibitem [{\citenamefont {Kane}(1998)}]{Kane1998}%
  \BibitemOpen
  \bibfield  {author} {\bibinfo {author} {\bibfnamefont {B.~E.}\ \bibnamefont
  {Kane}},\ }\bibfield  {title} {\enquote {\bibinfo {title} {{A silicon-based
  nuclear spin quantum computer}},}\ }\href {\doibase 10.1038/30156} {\bibfield
   {journal} {\bibinfo  {journal} {Nature}\ }\textbf {\bibinfo {volume}
  {393}},\ \bibinfo {pages} {133--137} (\bibinfo {year} {1998})}\BibitemShut
  {NoStop}%
\bibitem [{\citenamefont {Zwanenburg}\ \emph {et~al.}(2013)\citenamefont
  {Zwanenburg}, \citenamefont {Dzurak}, \citenamefont {Morello}, \citenamefont
  {Simmons}, \citenamefont {Hollenberg}, \citenamefont {Klimeck}, \citenamefont
  {Rogge}, \citenamefont {Coppersmith},\ and\ \citenamefont
  {Eriksson}}]{Zwanenburg2013}%
  \BibitemOpen
  \bibfield  {author} {\bibinfo {author} {\bibfnamefont {Floris~A.}\
  \bibnamefont {Zwanenburg}}, \bibinfo {author} {\bibfnamefont {Andrew~S.}\
  \bibnamefont {Dzurak}}, \bibinfo {author} {\bibfnamefont {Andrea}\
  \bibnamefont {Morello}}, \bibinfo {author} {\bibfnamefont {Michelle~Y.}\
  \bibnamefont {Simmons}}, \bibinfo {author} {\bibfnamefont {Lloyd C~L}\
  \bibnamefont {Hollenberg}}, \bibinfo {author} {\bibfnamefont {Gerhard}\
  \bibnamefont {Klimeck}}, \bibinfo {author} {\bibfnamefont {Sven}\
  \bibnamefont {Rogge}}, \bibinfo {author} {\bibfnamefont {Susan~N.}\
  \bibnamefont {Coppersmith}}, \ and\ \bibinfo {author} {\bibfnamefont
  {Mark~A.}\ \bibnamefont {Eriksson}},\ }\bibfield  {title} {\enquote {\bibinfo
  {title} {{Silicon quantum electronics}},}\ }\href {\doibase
  10.1103/RevModPhys.85.961} {\bibfield  {journal} {\bibinfo  {journal} {Rev.
  Mod. Phys.}\ }\textbf {\bibinfo {volume} {85}},\ \bibinfo {pages} {961--1019}
  (\bibinfo {year} {2013})}\BibitemShut {NoStop}%
\bibitem [{\citenamefont {Pla}\ \emph {et~al.}(2013)\citenamefont {Pla},
  \citenamefont {Tan}, \citenamefont {Dehollain}, \citenamefont {Lim},
  \citenamefont {Morton}, \citenamefont {Zwanenburg}, \citenamefont {Jamieson},
  \citenamefont {Dzurak},\ and\ \citenamefont {Morello}}]{Pla2013}%
  \BibitemOpen
  \bibfield  {author} {\bibinfo {author} {\bibfnamefont {Jarryd~J.}\
  \bibnamefont {Pla}}, \bibinfo {author} {\bibfnamefont {Kuan~Y.}\ \bibnamefont
  {Tan}}, \bibinfo {author} {\bibfnamefont {Juan~P.}\ \bibnamefont
  {Dehollain}}, \bibinfo {author} {\bibfnamefont {Wee~H.}\ \bibnamefont {Lim}},
  \bibinfo {author} {\bibfnamefont {John J.~L.}\ \bibnamefont {Morton}},
  \bibinfo {author} {\bibfnamefont {Floris~A.}\ \bibnamefont {Zwanenburg}},
  \bibinfo {author} {\bibfnamefont {David~N.}\ \bibnamefont {Jamieson}},
  \bibinfo {author} {\bibfnamefont {Andrew~S.}\ \bibnamefont {Dzurak}}, \ and\
  \bibinfo {author} {\bibfnamefont {Andrea}\ \bibnamefont {Morello}},\
  }\bibfield  {title} {\enquote {\bibinfo {title} {{High-fidelity readout and
  control of a nuclear spin qubit in silicon}},}\ }\href {\doibase
  10.1038/nature12011} {\bibfield  {journal} {\bibinfo  {journal} {Nature}\
  }\textbf {\bibinfo {volume} {496}},\ \bibinfo {pages} {334--338} (\bibinfo
  {year} {2013})}\BibitemShut {NoStop}%
\bibitem [{\citenamefont {Muhonen}\ \emph {et~al.}(2015)\citenamefont
  {Muhonen}, \citenamefont {Laucht}, \citenamefont {Simmons}, \citenamefont
  {Dehollain}, \citenamefont {Kalra}, \citenamefont {Hudson}, \citenamefont
  {Freer}, \citenamefont {Itoh}, \citenamefont {Jamieson}, \citenamefont
  {McCallum}, \citenamefont {Dzurak},\ and\ \citenamefont
  {Morello}}]{Muhonen2015}%
  \BibitemOpen
  \bibfield  {author} {\bibinfo {author} {\bibfnamefont {J~T}\ \bibnamefont
  {Muhonen}}, \bibinfo {author} {\bibfnamefont {A}~\bibnamefont {Laucht}},
  \bibinfo {author} {\bibfnamefont {S}~\bibnamefont {Simmons}}, \bibinfo
  {author} {\bibfnamefont {J~P}\ \bibnamefont {Dehollain}}, \bibinfo {author}
  {\bibfnamefont {R}~\bibnamefont {Kalra}}, \bibinfo {author} {\bibfnamefont
  {F~E}\ \bibnamefont {Hudson}}, \bibinfo {author} {\bibfnamefont
  {S}~\bibnamefont {Freer}}, \bibinfo {author} {\bibfnamefont {K~M}\
  \bibnamefont {Itoh}}, \bibinfo {author} {\bibfnamefont {D~N}\ \bibnamefont
  {Jamieson}}, \bibinfo {author} {\bibfnamefont {J~C}\ \bibnamefont
  {McCallum}}, \bibinfo {author} {\bibfnamefont {A~S}\ \bibnamefont {Dzurak}},
  \ and\ \bibinfo {author} {\bibfnamefont {A}~\bibnamefont {Morello}},\
  }\bibfield  {title} {\enquote {\bibinfo {title} {{Quantifying the quantum
  gate fidelity of single-atom spin qubits in silicon by randomized
  benchmarking}},}\ }\href {\doibase 10.1088/0953-8984/27/15/154205} {\bibfield
   {journal} {\bibinfo  {journal} {J. Phys. Condens. Matter}\ }\textbf
  {\bibinfo {volume} {27}},\ \bibinfo {pages} {154205} (\bibinfo {year}
  {2015})}\BibitemShut {NoStop}%
\bibitem [{\citenamefont {Muhonen}\ \emph {et~al.}(2017)\citenamefont
  {Muhonen}, \citenamefont {Dehollain}, \citenamefont {Laucht}, \citenamefont
  {Simmons}, \citenamefont {Kalra}, \citenamefont {Hudson}, \citenamefont
  {Jamieson}, \citenamefont {McCallum}, \citenamefont {Itoh}, \citenamefont
  {Dzurak},\ and\ \citenamefont {Morello}}]{Muhonen2018}%
  \BibitemOpen
  \bibfield  {author} {\bibinfo {author} {\bibfnamefont {J.~T.}\ \bibnamefont
  {Muhonen}}, \bibinfo {author} {\bibfnamefont {J.~P.}\ \bibnamefont
  {Dehollain}}, \bibinfo {author} {\bibfnamefont {A.}~\bibnamefont {Laucht}},
  \bibinfo {author} {\bibfnamefont {S.}~\bibnamefont {Simmons}}, \bibinfo
  {author} {\bibfnamefont {R.}~\bibnamefont {Kalra}}, \bibinfo {author}
  {\bibfnamefont {F.~E.}\ \bibnamefont {Hudson}}, \bibinfo {author}
  {\bibfnamefont {D.~N.}\ \bibnamefont {Jamieson}}, \bibinfo {author}
  {\bibfnamefont {J.~C.}\ \bibnamefont {McCallum}}, \bibinfo {author}
  {\bibfnamefont {K.~M.}\ \bibnamefont {Itoh}}, \bibinfo {author}
  {\bibfnamefont {A.~S.}\ \bibnamefont {Dzurak}}, \ and\ \bibinfo {author}
  {\bibfnamefont {A.}~\bibnamefont {Morello}},\ }\bibfield  {title} {\enquote
  {\bibinfo {title} {{Coherent control via weak measurements in $^{31}$P
  single-atom electron and nuclear spin qubits}},}\ }\href {\doibase
  10.1103/PhysRevB.98.155201} {\bibfield  {journal} {\bibinfo  {journal} {Phys.
  Rev. B}\ }\textbf {\bibinfo {volume} {98}},\ \bibinfo {pages} {155201}
  (\bibinfo {year} {2017})}\BibitemShut {NoStop}%
\bibitem [{\citenamefont {Itoh}\ and\ \citenamefont
  {Watanabe}(2014)}]{Itoh2014}%
  \BibitemOpen
  \bibfield  {author} {\bibinfo {author} {\bibfnamefont {Kohei~M.}\
  \bibnamefont {Itoh}}\ and\ \bibinfo {author} {\bibfnamefont {Hideyuki}\
  \bibnamefont {Watanabe}},\ }\bibfield  {title} {\enquote {\bibinfo {title}
  {{Isotope engineering of silicon and diamond for quantum computing and
  sensing applications}},}\ }\href {\doibase 10.1557/mrc.2014.32} {\bibfield
  {journal} {\bibinfo  {journal} {MRS Commun.}\ }\textbf {\bibinfo {volume}
  {4}},\ \bibinfo {pages} {143--157} (\bibinfo {year} {2014})}\BibitemShut
  {NoStop}%
\bibitem [{\citenamefont {Sigillito}\ \emph {et~al.}(2017)\citenamefont
  {Sigillito}, \citenamefont {Tyryshkin}, \citenamefont {Schenkel},
  \citenamefont {Houck},\ and\ \citenamefont {Lyon}}]{Sigillito2017b}%
  \BibitemOpen
  \bibfield  {author} {\bibinfo {author} {\bibfnamefont {Anthony~J.}\
  \bibnamefont {Sigillito}}, \bibinfo {author} {\bibfnamefont {Alexei~M.}\
  \bibnamefont {Tyryshkin}}, \bibinfo {author} {\bibfnamefont {Thomas}\
  \bibnamefont {Schenkel}}, \bibinfo {author} {\bibfnamefont {Andrew~A.}\
  \bibnamefont {Houck}}, \ and\ \bibinfo {author} {\bibfnamefont {Stephen~A.}\
  \bibnamefont {Lyon}},\ }\bibfield  {title} {\enquote {\bibinfo {title}
  {{All-electric control of donor nuclear spin qubits in silicon}},}\ }\href
  {\doibase 10.1038/nnano.2017.154} {\bibfield  {journal} {\bibinfo  {journal}
  {Nat. Nanotechnol.}\ }\textbf {\bibinfo {volume} {12}},\ \bibinfo {pages}
  {958--962} (\bibinfo {year} {2017})}\BibitemShut {NoStop}%
\bibitem [{\citenamefont {Hill}\ \emph {et~al.}(2015)\citenamefont {Hill},
  \citenamefont {Peretz}, \citenamefont {Hile}, \citenamefont {House},
  \citenamefont {Fuechsle}, \citenamefont {Rogge}, \citenamefont {Simmons},\
  and\ \citenamefont {Hollenberg}}]{Hill2015}%
  \BibitemOpen
  \bibfield  {author} {\bibinfo {author} {\bibfnamefont {Charles~D.}\
  \bibnamefont {Hill}}, \bibinfo {author} {\bibfnamefont {Eldad}\ \bibnamefont
  {Peretz}}, \bibinfo {author} {\bibfnamefont {Samuel~J.}\ \bibnamefont
  {Hile}}, \bibinfo {author} {\bibfnamefont {Matthew~G.}\ \bibnamefont
  {House}}, \bibinfo {author} {\bibfnamefont {Martin}\ \bibnamefont
  {Fuechsle}}, \bibinfo {author} {\bibfnamefont {Sven}\ \bibnamefont {Rogge}},
  \bibinfo {author} {\bibfnamefont {Michelle~Y.}\ \bibnamefont {Simmons}}, \
  and\ \bibinfo {author} {\bibfnamefont {Lloyd C.L.~L.}\ \bibnamefont
  {Hollenberg}},\ }\bibfield  {title} {\enquote {\bibinfo {title} {{A surface
  code quantum computer in silicon}},}\ }\href {\doibase
  10.1126/sciadv.1500707} {\bibfield  {journal} {\bibinfo  {journal} {Sci.
  Adv.}\ }\textbf {\bibinfo {volume} {1}},\ \bibinfo {pages} {e1500707}
  (\bibinfo {year} {2015})}\BibitemShut {NoStop}%
\bibitem [{\citenamefont {Hile}\ \emph {et~al.}(2018)\citenamefont {Hile},
  \citenamefont {Fricke}, \citenamefont {House}, \citenamefont {Peretz},
  \citenamefont {Chen}, \citenamefont {Wang}, \citenamefont {Broome},
  \citenamefont {Gorman}, \citenamefont {Keizer}, \citenamefont {Rahman},\ and\
  \citenamefont {Simmons}}]{Hile2018}%
  \BibitemOpen
  \bibfield  {author} {\bibinfo {author} {\bibfnamefont {Samuel~J.}\
  \bibnamefont {Hile}}, \bibinfo {author} {\bibfnamefont {Lukas}\ \bibnamefont
  {Fricke}}, \bibinfo {author} {\bibfnamefont {Matthew~G.}\ \bibnamefont
  {House}}, \bibinfo {author} {\bibfnamefont {Eldad}\ \bibnamefont {Peretz}},
  \bibinfo {author} {\bibfnamefont {Chin~Yi}\ \bibnamefont {Chen}}, \bibinfo
  {author} {\bibfnamefont {Yu}~\bibnamefont {Wang}}, \bibinfo {author}
  {\bibfnamefont {Matthew}\ \bibnamefont {Broome}}, \bibinfo {author}
  {\bibfnamefont {Samuel~K.}\ \bibnamefont {Gorman}}, \bibinfo {author}
  {\bibfnamefont {Joris~G.}\ \bibnamefont {Keizer}}, \bibinfo {author}
  {\bibfnamefont {Rajib}\ \bibnamefont {Rahman}}, \ and\ \bibinfo {author}
  {\bibfnamefont {Michelle~Y.}\ \bibnamefont {Simmons}},\ }\bibfield  {title}
  {\enquote {\bibinfo {title} {Addressable electron spin resonance using donors
  and donor molecules in silicon},}\ }\href {\doibase 10.1126/sciadv.aaq1459}
  {\bibfield  {journal} {\bibinfo  {journal} {Science Advances}\ }\textbf
  {\bibinfo {volume} {4}} (\bibinfo {year} {2018}),\
  10.1126/sciadv.aaq1459}\BibitemShut {NoStop}%
\bibitem [{\citenamefont {He}\ \emph {et~al.}(2019)\citenamefont {He},
  \citenamefont {Gorman}, \citenamefont {Keith}, \citenamefont {Kranz},
  \citenamefont {Keizer},\ and\ \citenamefont {Simmons}}]{He2019}%
  \BibitemOpen
  \bibfield  {author} {\bibinfo {author} {\bibfnamefont {Y.}~\bibnamefont
  {He}}, \bibinfo {author} {\bibfnamefont {S.~K.}\ \bibnamefont {Gorman}},
  \bibinfo {author} {\bibfnamefont {D.}~\bibnamefont {Keith}}, \bibinfo
  {author} {\bibfnamefont {L.}~\bibnamefont {Kranz}}, \bibinfo {author}
  {\bibfnamefont {J.~G.}\ \bibnamefont {Keizer}}, \ and\ \bibinfo {author}
  {\bibfnamefont {M.~Y.}\ \bibnamefont {Simmons}},\ }\bibfield  {title}
  {\enquote {\bibinfo {title} {A two-qubit gate between phosphorus donor
  electrons in silicon},}\ }\href {\doibase 10.1038/s41586-019-1381-2}
  {\bibfield  {journal} {\bibinfo  {journal} {Nature}\ }\textbf {\bibinfo
  {volume} {571}},\ \bibinfo {pages} {371--375} (\bibinfo {year}
  {2019})}\BibitemShut {NoStop}%
\bibitem [{\citenamefont {Tosi}\ \emph {et~al.}(2018)\citenamefont {Tosi},
  \citenamefont {Mohiyaddin}, \citenamefont {Tenberg}, \citenamefont {Laucht},\
  and\ \citenamefont {Morello}}]{Tosi2017}%
  \BibitemOpen
  \bibfield  {author} {\bibinfo {author} {\bibfnamefont {Guilherme}\
  \bibnamefont {Tosi}}, \bibinfo {author} {\bibfnamefont {Fahd~A.}\
  \bibnamefont {Mohiyaddin}}, \bibinfo {author} {\bibfnamefont {Stefanie}\
  \bibnamefont {Tenberg}}, \bibinfo {author} {\bibfnamefont {Arne}\
  \bibnamefont {Laucht}}, \ and\ \bibinfo {author} {\bibfnamefont {Andrea}\
  \bibnamefont {Morello}},\ }\bibfield  {title} {\enquote {\bibinfo {title}
  {{Robust electric dipole transition at microwave frequencies for nuclear spin
  qubits in silicon}},}\ }\href {\doibase 10.1103/PhysRevB.98.075313}
  {\bibfield  {journal} {\bibinfo  {journal} {Phys. Rev. B}\ }\textbf {\bibinfo
  {volume} {98}},\ \bibinfo {pages} {075313} (\bibinfo {year}
  {2018})}\BibitemShut {NoStop}%
\bibitem [{\citenamefont {Tosi}\ \emph {et~al.}(2017)\citenamefont {Tosi},
  \citenamefont {Mohiyaddin}, \citenamefont {Schmitt}, \citenamefont {Tenberg},
  \citenamefont {Rahman}, \citenamefont {Klimeck},\ and\ \citenamefont
  {Morello}}]{Tosi2017b}%
  \BibitemOpen
  \bibfield  {author} {\bibinfo {author} {\bibfnamefont {Guilherme}\
  \bibnamefont {Tosi}}, \bibinfo {author} {\bibfnamefont {Fahd~A.}\
  \bibnamefont {Mohiyaddin}}, \bibinfo {author} {\bibfnamefont {Vivien}\
  \bibnamefont {Schmitt}}, \bibinfo {author} {\bibfnamefont {Stefanie}\
  \bibnamefont {Tenberg}}, \bibinfo {author} {\bibfnamefont {Rajib}\
  \bibnamefont {Rahman}}, \bibinfo {author} {\bibfnamefont {Gerhard}\
  \bibnamefont {Klimeck}}, \ and\ \bibinfo {author} {\bibfnamefont {Andrea}\
  \bibnamefont {Morello}},\ }\bibfield  {title} {\enquote {\bibinfo {title}
  {{Silicon quantum processor with robust long-distance qubit couplings}},}\
  }\href {\doibase 10.1038/s41467-017-00378-x} {\bibfield  {journal} {\bibinfo
  {journal} {Nat. Commun.}\ }\textbf {\bibinfo {volume} {8}},\ \bibinfo {pages}
  {450} (\bibinfo {year} {2017})}\BibitemShut {NoStop}%
\bibitem [{\citenamefont {Freeman}\ \emph {et~al.}(2016)\citenamefont
  {Freeman}, \citenamefont {Schoenfield},\ and\ \citenamefont
  {Jiang}}]{Freeman2016}%
  \BibitemOpen
  \bibfield  {author} {\bibinfo {author} {\bibfnamefont {Blake~M.}\
  \bibnamefont {Freeman}}, \bibinfo {author} {\bibfnamefont {Joshua~S.}\
  \bibnamefont {Schoenfield}}, \ and\ \bibinfo {author} {\bibfnamefont
  {Hongwen}\ \bibnamefont {Jiang}},\ }\bibfield  {title} {\enquote {\bibinfo
  {title} {{Comparison of low frequency charge noise in identically patterned
  Si/SiO 2 and Si/SiGe quantum dots}},}\ }\href {\doibase 10.1063/1.4954700}
  {\bibfield  {journal} {\bibinfo  {journal} {Appl. Phys. Lett.}\ }\textbf
  {\bibinfo {volume} {108}},\ \bibinfo {pages} {253108} (\bibinfo {year}
  {2016})}\BibitemShut {NoStop}%
\bibitem [{\citenamefont {Weber}\ \emph {et~al.}(2018)\citenamefont {Weber},
  \citenamefont {Hsueh}, \citenamefont {Watson}, \citenamefont {Li},
  \citenamefont {Hamilton}, \citenamefont {Hollenberg}, \citenamefont
  {Rahman},\ and\ \citenamefont {Simmons}}]{Weber2018}%
  \BibitemOpen
  \bibfield  {author} {\bibinfo {author} {\bibfnamefont {Bent}\ \bibnamefont
  {Weber}}, \bibinfo {author} {\bibfnamefont {Yu-Ling}\ \bibnamefont {Hsueh}},
  \bibinfo {author} {\bibfnamefont {Thomas~F.}\ \bibnamefont {Watson}},
  \bibinfo {author} {\bibfnamefont {Ruoyu}\ \bibnamefont {Li}}, \bibinfo
  {author} {\bibfnamefont {Alexander~R.}\ \bibnamefont {Hamilton}}, \bibinfo
  {author} {\bibfnamefont {Lloyd C.~L.}\ \bibnamefont {Hollenberg}}, \bibinfo
  {author} {\bibfnamefont {Rajib}\ \bibnamefont {Rahman}}, \ and\ \bibinfo
  {author} {\bibfnamefont {Michelle~Y.}\ \bibnamefont {Simmons}},\ }\bibfield
  {title} {\enquote {\bibinfo {title} {Spin–orbit coupling in silicon for
  electrons bound to donors},}\ }\href {\doibase 10.1038/s41534-018-0111-1}
  {\bibfield  {journal} {\bibinfo  {journal} {npj Quantum Information}\
  }\textbf {\bibinfo {volume} {4}} (\bibinfo {year} {2018}),\
  10.1038/s41534-018-0111-1}\BibitemShut {NoStop}%
\bibitem [{\citenamefont {Boross}\ \emph {et~al.}(2018)\citenamefont {Boross},
  \citenamefont {Sz\'{e}chenyi},\ and\ \citenamefont {P\'{a}lyi}}]{Boross2018}%
  \BibitemOpen
  \bibfield  {author} {\bibinfo {author} {\bibfnamefont {P\'{e}ter}\
  \bibnamefont {Boross}}, \bibinfo {author} {\bibfnamefont {G\'{a}bor}\
  \bibnamefont {Sz\'{e}chenyi}}, \ and\ \bibinfo {author} {\bibfnamefont
  {Andr\'{a}s}\ \bibnamefont {P\'{a}lyi}},\ }\bibfield  {title} {\enquote
  {\bibinfo {title} {Hyperfine-assisted fast electric control of dopant nuclear
  spins in semiconductors},}\ }\href {\doibase 10.1103/physrevb.97.245417}
  {\bibfield  {journal} {\bibinfo  {journal} {Physical Review B}\ }\textbf
  {\bibinfo {volume} {97}} (\bibinfo {year} {2018}),\
  10.1103/physrevb.97.245417}\BibitemShut {NoStop}%
\bibitem [{\citenamefont {Bloch}\ and\ \citenamefont
  {Siegert}(1940)}]{BlochSiegert}%
  \BibitemOpen
  \bibfield  {author} {\bibinfo {author} {\bibfnamefont {F.}~\bibnamefont
  {Bloch}}\ and\ \bibinfo {author} {\bibfnamefont {A.}~\bibnamefont
  {Siegert}},\ }\bibfield  {title} {\enquote {\bibinfo {title} {{Magnetic
  Resonance for Nonrotating Fields}},}\ }\href {\doibase
  10.1103/PhysRev.57.522} {\bibfield  {journal} {\bibinfo  {journal} {Phys.
  Rev.}\ }\textbf {\bibinfo {volume} {57}},\ \bibinfo {pages} {522--527}
  (\bibinfo {year} {1940})}\BibitemShut {NoStop}%
\bibitem [{\citenamefont {Paladino}\ \emph {et~al.}(2013)\citenamefont
  {Paladino}, \citenamefont {Galperin}, \citenamefont {Falci},\ and\
  \citenamefont {Altshuler}}]{Paladino2014}%
  \BibitemOpen
  \bibfield  {author} {\bibinfo {author} {\bibfnamefont {E.}~\bibnamefont
  {Paladino}}, \bibinfo {author} {\bibfnamefont {Y.~M.}\ \bibnamefont
  {Galperin}}, \bibinfo {author} {\bibfnamefont {G.}~\bibnamefont {Falci}}, \
  and\ \bibinfo {author} {\bibfnamefont {B.~L.}\ \bibnamefont {Altshuler}},\
  }\bibfield  {title} {\enquote {\bibinfo {title} {1/f noise: implications for
  solid-state quantum information},}\ }\href {\doibase
  10.1103/RevModPhys.86.361} {\bibfield  {journal} {\bibinfo  {journal} {Rev.
  Mod. Phys.}\ }\textbf {\bibinfo {volume} {86}},\ \bibinfo {pages} {361--418}
  (\bibinfo {year} {2013})}\BibitemShut {NoStop}%
\bibitem [{Not()}]{Note_Lateral_Noise}%
  \BibitemOpen
  \href@noop {} {}\bibinfo {note} {Horizontal noise acting on the tunnel
  coupling $V_t$, stemming from the tuning of the tunnel coupling as proposed
  by Ref.~\onlinecite{Tosi2017b}, would cause negligible errors as also shown
  by Ref.~\onlinecite{Tosi2017b})}\BibitemShut {NoStop}%
\bibitem [{\citenamefont {Wimperis}(1994)}]{Wimperis1994}%
  \BibitemOpen
  \bibfield  {author} {\bibinfo {author} {\bibfnamefont {S.}~\bibnamefont
  {Wimperis}},\ }\bibfield  {title} {\enquote {\bibinfo {title} {Broadband,
  narrowband, and passband composite pulses for use in advanced nmr
  experiments},}\ }\href {\doibase https://doi.org/10.1006/jmra.1994.1159}
  {\bibfield  {journal} {\bibinfo  {journal} {Journal of Magnetic Resonance,
  Series A}\ }\textbf {\bibinfo {volume} {109}},\ \bibinfo {pages} {221 -- 231}
  (\bibinfo {year} {1994})}\BibitemShut {NoStop}%
\bibitem [{\citenamefont {Harvey-Collard}\ \emph {et~al.}(2017)\citenamefont
  {Harvey-Collard}, \citenamefont {Jacobson}, \citenamefont {Rudolph},
  \citenamefont {Dominguez}, \citenamefont {{Ten Eyck}}, \citenamefont {Wendt},
  \citenamefont {Pluym}, \citenamefont {Gamble}, \citenamefont {Lilly},
  \citenamefont {Pioro-Ladri{\`{e}}re},\ and\ \citenamefont
  {Carroll}}]{Harvey-Collard2017a}%
  \BibitemOpen
  \bibfield  {author} {\bibinfo {author} {\bibfnamefont {Patrick}\ \bibnamefont
  {Harvey-Collard}}, \bibinfo {author} {\bibfnamefont {N.~Tobias}\ \bibnamefont
  {Jacobson}}, \bibinfo {author} {\bibfnamefont {Martin}\ \bibnamefont
  {Rudolph}}, \bibinfo {author} {\bibfnamefont {Jason}\ \bibnamefont
  {Dominguez}}, \bibinfo {author} {\bibfnamefont {Gregory~A.}\ \bibnamefont
  {{Ten Eyck}}}, \bibinfo {author} {\bibfnamefont {Joel~R.}\ \bibnamefont
  {Wendt}}, \bibinfo {author} {\bibfnamefont {Tammy}\ \bibnamefont {Pluym}},
  \bibinfo {author} {\bibfnamefont {John~King}\ \bibnamefont {Gamble}},
  \bibinfo {author} {\bibfnamefont {Michael~P.}\ \bibnamefont {Lilly}},
  \bibinfo {author} {\bibfnamefont {Michel}\ \bibnamefont
  {Pioro-Ladri{\`{e}}re}}, \ and\ \bibinfo {author} {\bibfnamefont
  {Malcolm~S.}\ \bibnamefont {Carroll}},\ }\bibfield  {title} {\enquote
  {\bibinfo {title} {{Coherent coupling between a quantum dot and a donor in
  silicon}},}\ }\href {\doibase 10.1038/s41467-017-01113-2} {\bibfield
  {journal} {\bibinfo  {journal} {Nat. Commun.}\ }\textbf {\bibinfo {volume}
  {8}},\ \bibinfo {pages} {1029} (\bibinfo {year} {2017})}\BibitemShut
  {NoStop}%
\bibitem [{\citenamefont {Martins}\ \emph {et~al.}(2016)\citenamefont
  {Martins}, \citenamefont {Malinowski}, \citenamefont {Nissen}, \citenamefont
  {Barnes}, \citenamefont {Fallahi}, \citenamefont {Gardner}, \citenamefont
  {Manfra}, \citenamefont {Marcus},\ and\ \citenamefont
  {Kuemmeth}}]{Martins2016}%
  \BibitemOpen
  \bibfield  {author} {\bibinfo {author} {\bibfnamefont {Frederico}\
  \bibnamefont {Martins}}, \bibinfo {author} {\bibfnamefont {Filip~K.}\
  \bibnamefont {Malinowski}}, \bibinfo {author} {\bibfnamefont {Peter~D.}\
  \bibnamefont {Nissen}}, \bibinfo {author} {\bibfnamefont {Edwin}\
  \bibnamefont {Barnes}}, \bibinfo {author} {\bibfnamefont {Saeed}\
  \bibnamefont {Fallahi}}, \bibinfo {author} {\bibfnamefont {Geoffrey~C.}\
  \bibnamefont {Gardner}}, \bibinfo {author} {\bibfnamefont {Michael~J.}\
  \bibnamefont {Manfra}}, \bibinfo {author} {\bibfnamefont {Charles~M.}\
  \bibnamefont {Marcus}}, \ and\ \bibinfo {author} {\bibfnamefont {Ferdinand}\
  \bibnamefont {Kuemmeth}},\ }\bibfield  {title} {\enquote {\bibinfo {title}
  {Noise suppression using symmetric exchange gates in spin qubits},}\ }\href
  {\doibase 10.1103/PhysRevLett.116.116801} {\bibfield  {journal} {\bibinfo
  {journal} {Phys. Rev. Lett.}\ }\textbf {\bibinfo {volume} {116}},\ \bibinfo
  {pages} {116801} (\bibinfo {year} {2016})}\BibitemShut {NoStop}%
\bibitem [{\citenamefont {Pedersen}\ \emph {et~al.}(2007)\citenamefont
  {Pedersen}, \citenamefont {M{\o}ller},\ and\ \citenamefont
  {M{\o}lmer}}]{Pedersen2007}%
  \BibitemOpen
  \bibfield  {author} {\bibinfo {author} {\bibfnamefont {Line~Hjortsh{\o}j}\
  \bibnamefont {Pedersen}}, \bibinfo {author} {\bibfnamefont {Niels~Martin}\
  \bibnamefont {M{\o}ller}}, \ and\ \bibinfo {author} {\bibfnamefont {Klaus}\
  \bibnamefont {M{\o}lmer}},\ }\bibfield  {title} {\enquote {\bibinfo {title}
  {{Fidelity of quantum operations}},}\ }\href {\doibase
  10.1016/j.physleta.2007.02.069} {\bibfield  {journal} {\bibinfo  {journal}
  {Phys. Lett. A}\ }\textbf {\bibinfo {volume} {367}},\ \bibinfo {pages}
  {47--51} (\bibinfo {year} {2007})}\BibitemShut {NoStop}%
\bibitem [{\citenamefont {Kok}\ and\ \citenamefont
  {Lovett}(2010)}]{RamanTransition}%
  \BibitemOpen
  \bibfield  {author} {\bibinfo {author} {\bibfnamefont {P.}~\bibnamefont
  {Kok}}\ and\ \bibinfo {author} {\bibfnamefont {B.~W.}\ \bibnamefont
  {Lovett}},\ }\href@noop {} {\emph {\bibinfo {title} {Introduction to Quantum
  Optical Information Processing}}}\ (\bibinfo  {publisher} {Cambridge
  University Press},\ \bibinfo {year} {2010})\ Chap.\ \bibinfo {chapter}
  {Section 7.1.3}\BibitemShut {NoStop}%
\bibitem [{\citenamefont {Garwood}\ and\ \citenamefont
  {Delabarre}(2001)}]{Garwood2001}%
  \BibitemOpen
  \bibfield  {author} {\bibinfo {author} {\bibfnamefont {Michael}\ \bibnamefont
  {Garwood}}\ and\ \bibinfo {author} {\bibfnamefont {Lance}\ \bibnamefont
  {Delabarre}},\ }\bibfield  {title} {\enquote {\bibinfo {title} {The return of
  the frequency sweep: Designing adiabatic pulses for contemporary nmr},}\
  }\href {\doibase 10.1006/jmre.2001.2340} {\bibfield  {journal} {\bibinfo
  {journal} {Journal of Magnetic Resonance}\ }\textbf {\bibinfo {volume}
  {153}},\ \bibinfo {pages} {155–177} (\bibinfo {year} {2001})}\BibitemShut
  {NoStop}%
\bibitem [{\citenamefont {Fowler}\ \emph {et~al.}(2012)\citenamefont {Fowler},
  \citenamefont {Whiteside},\ and\ \citenamefont {Hollenberg}}]{Fowler2012a}%
  \BibitemOpen
  \bibfield  {author} {\bibinfo {author} {\bibfnamefont {Austin~G.}\
  \bibnamefont {Fowler}}, \bibinfo {author} {\bibfnamefont {Adam~C.}\
  \bibnamefont {Whiteside}}, \ and\ \bibinfo {author} {\bibfnamefont {Lloyd
  C~L}\ \bibnamefont {Hollenberg}},\ }\bibfield  {title} {\enquote {\bibinfo
  {title} {{Towards practical classical processing for the surface code: Timing
  analysis}},}\ }\href {\doibase 10.1103/PhysRevA.86.042313} {\bibfield
  {journal} {\bibinfo  {journal} {Phys. Rev. A}\ }\textbf {\bibinfo {volume}
  {86}},\ \bibinfo {pages} {042313} (\bibinfo {year} {2012})}\BibitemShut
  {NoStop}%
\bibitem [{\citenamefont {Raussendorf}\ and\ \citenamefont
  {Harrington}(2007)}]{Raussendorf2007}%
  \BibitemOpen
  \bibfield  {author} {\bibinfo {author} {\bibfnamefont {Robert}\ \bibnamefont
  {Raussendorf}}\ and\ \bibinfo {author} {\bibfnamefont {Jim}\ \bibnamefont
  {Harrington}},\ }\bibfield  {title} {\enquote {\bibinfo {title}
  {{Fault-Tolerant Quantum Computation with High Threshold in Two
  Dimensions}},}\ }\href {\doibase 10.1103/PhysRevLett.98.190504} {\bibfield
  {journal} {\bibinfo  {journal} {Phys. Rev. Lett.}\ }\textbf {\bibinfo
  {volume} {98}},\ \bibinfo {pages} {190504} (\bibinfo {year}
  {2007})}\BibitemShut {NoStop}%
\bibitem [{\citenamefont {Shirley}(1965)}]{floquetTheory}%
  \BibitemOpen
  \bibfield  {author} {\bibinfo {author} {\bibfnamefont {Jon~H.}\ \bibnamefont
  {Shirley}},\ }\bibfield  {title} {\enquote {\bibinfo {title} {{Solution of
  the Schr{\"{o}}dinger Equation with a Hamiltonian Periodic in Time}},}\
  }\href {\doibase 10.1103/PhysRev.138.B979} {\bibfield  {journal} {\bibinfo
  {journal} {Phys. Rev.}\ }\textbf {\bibinfo {volume} {138}},\ \bibinfo {pages}
  {B979--B987} (\bibinfo {year} {1965})}\BibitemShut {NoStop}%
\bibitem [{\citenamefont {Ho}\ \emph {et~al.}(1983)\citenamefont {Ho},
  \citenamefont {Chu},\ and\ \citenamefont {Tietz}}]{multiFreq}%
  \BibitemOpen
  \bibfield  {author} {\bibinfo {author} {\bibfnamefont {Tak-San}\ \bibnamefont
  {Ho}}, \bibinfo {author} {\bibfnamefont {Shih-I}\ \bibnamefont {Chu}}, \ and\
  \bibinfo {author} {\bibfnamefont {James~V.}\ \bibnamefont {Tietz}},\
  }\bibfield  {title} {\enquote {\bibinfo {title} {{Semiclassical many-mode
  floquet theory}},}\ }\href {\doibase 10.1016/0009-2614(83)80732-5} {\bibfield
   {journal} {\bibinfo  {journal} {Chem. Phys. Lett.}\ }\textbf {\bibinfo
  {volume} {96}},\ \bibinfo {pages} {464--471} (\bibinfo {year}
  {1983})}\BibitemShut {NoStop}%
\bibitem [{\citenamefont {Winkler}(2003)}]{schriefferWolff}%
  \BibitemOpen
  \bibfield  {author} {\bibinfo {author} {\bibfnamefont {R.}~\bibnamefont
  {Winkler}},\ }\href@noop {} {\emph {\bibinfo {title} {Spin-orbit Coupling
  Effects in Two-Dimensional Electron and Hole Systems}}}\ (\bibinfo
  {publisher} {Springer},\ \bibinfo {year} {2003})\ Chap.\ \bibinfo {chapter}
  {Appendix B}\BibitemShut {NoStop}%
\end{thebibliography}%
\end{document}